\documentclass{ieeeaccess}

\usepackage{url}
\usepackage{cite}
\usepackage{amsmath,amssymb,amsfonts}
\usepackage{algorithmic}
\usepackage{graphicx}
\usepackage[export]{adjustbox}
\usepackage{textcomp}
\usepackage{comment}
\usepackage{stfloats}
\graphicspath{{../figs/}}
\usepackage{caption}
\usepackage{listings}
\usepackage{subcaption}
\DeclareGraphicsExtensions{.png}
\usepackage{float}
\def\BibTeX{{\rm B\kern-.05em{\sc i\kern-.025em b}\kern-.08em
    T\kern-.1667em\lower.7ex\hbox{E}\kern-.125emX}}

\begin{document}
\history{Date of publication xxxx 00, 0000, date of current version xxxx 00, 0000.}
\doi{10.1109/ACCESS.2017.DOI}

\title{Can we run our Ethereum nodes at home?}

\author{
    \uppercase{Mikel Cortes-Goicoechea}\authorrefmark{1},
    \uppercase{Tarun Mohandas-Daryanani}\authorrefmark{1}, 
    \uppercase{Jose Luis Muñoz-Tapia}\authorrefmark{3}, 
    \uppercase{and Leonardo Bautista-Gomez}.\authorrefmark{4}
}
\address[1]{Barcelona Supercomputing Center, Barcelona, Spain (e-mail: mikel.cortes@bsc.es, e-mail: tarun.mohandas@bsc.es)}
\address[3]{Universidad Politécnica de Catalunya, Barcelona, Spain (e-mail: jose.luis.munoz@upc.edu)}
\address[3]{Status.im \& MigaLabs, Barcelona, Spain (e-mail: leo@status.im)}


\markboth
{Cortes \headeretal: Preparation of Papers for IEEE TRANSACTIONS and JOURNALS}
{Cortes \headeretal: Preparation of Papers for IEEE TRANSACTIONS and JOURNALS}

\corresp{Corresponding author: Mikel Cortes-Goicoechea (e-mail: mikel.cortes@bsc.es).}


\begin{abstract}
Scalability is a common issue among the most used permissionless blockchains, and several approaches have been proposed to solve this issue. Tackling scalability while preserving the security and decentralization of the network is a significant challenge. To deliver effective scaling solutions, Ethereum achieved a major protocol improvement, including a change in the consensus mechanism towards Proof of Stake. This improvement aimed a vast reduction of the hardware requirements to run a node, leading to significant sustainability benefits with a lower network energy consumption. This work analyzes the resource usage behavior of different clients running as Ethereum consensus nodes, comparing their performance under different configurations and analyzing their differences. Our results show higher requirements than claimed initially and how different clients react to network perturbations. Furthermore, we discuss the differences between the consensus clients, including their strong points and limitations.
\end{abstract}


\titlepgskip=-15pt

\maketitle

\section{Introduction}
\label{sec:introduction}

Ethereum~\cite{eth-whitepaper} has been a remarkable achievement on the road to ubiquitous blockchain technology. It led to considerable growth in decentralized applications and the Web3 space due to its pioneer multipurpose Ethereum Virtual Machine (EVM)~\cite{hildenbrandt2018kevm} and its dedicated programming language Solidity~\cite{wohrer2018smart}. With an extended network of more than $10.000$ active nodes~\cite{cortes2021discovering} and its capabilities to process \emph{Smart Contracts}, it currently handles above $1$ million transactions per day from $600.000$ active accounts~\cite{oliveira2022analysis}. These characteristics have set the conditions for a solid community of developers and continuous advancements, as well as introducing new technological possibilities. As the Ethereum adoption increases, its usability has been threatened by the rising transaction volume and network clogging.

Following the highly tested consensus mechanism at the moment, initially, Ethereum relied on a Proof of Work (PoW) to reach consensus ~\cite{tikhomirov2018ethereum} among its participants. 
However, as the very first blockchain ever presented to the world, Bitcoin~\cite{nakamoto2009bitcoin}, Ethereum shared a similar set of limitations in terms of scalability and sustainability \cite{hayes2015cost}.

In an attempt to make the network more sustainable while presenting a solid foundation to scale up the network capabilities, in 2020, Ethereum embraced a live consensus transition towards a more sustainable Proof of Stake (PoS) \cite{saleh2021blockchain} based on Casper FFG for finalization \cite{buterin2020combining} and LMD GHOST \cite{Moindrot2017ProofOS} as a fork-choice rule.
With this transition, network participants no longer had to compete with each other to be the first ones to find a block with a valid hash, as the block proposers are randomly selected from the set of active validators using the RANDAO \cite{randao-code} algorithm.

\begin{figure*}[!htb]\hfill
    \includegraphics[width=\linewidth]{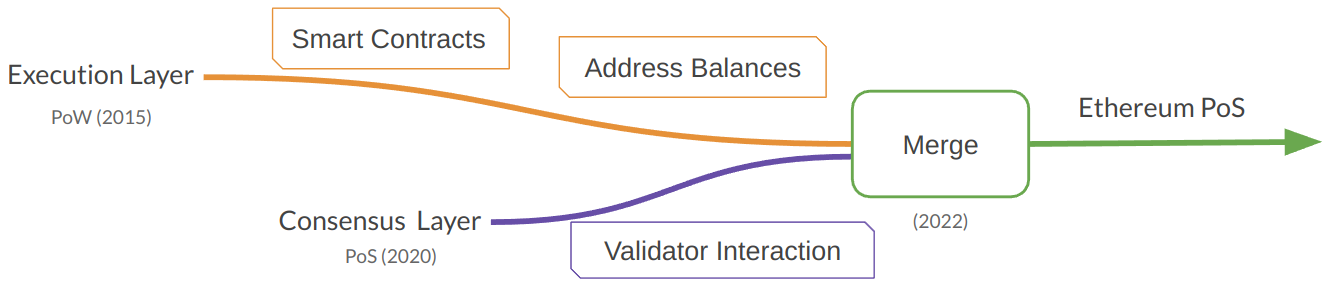}
    \caption{Ethereum transition from PoW to PoS.}
    \label{fig:eth-transition}
\end{figure*}

Due to the complexity of introducing this consensus mechanism, the transition was divided into different phases, where all the changes were gradually applied and tested before the final merge happened.
In the first phase, in December 2020, PoS was implemented on the Beacon Chain, a live and independent blockchain running in parallel to Ethereum's PoW main network. 
To participate in Ethereum's PoS consensus as a validator, users or entities must deposit the Ethereum PoS's smart contract \cite{park2020end}, activating new validators into the chain.

\begin{figure*}[!htb]
    \centering
    \minipage{0.45\textwidth}
        \includegraphics[width=\linewidth]{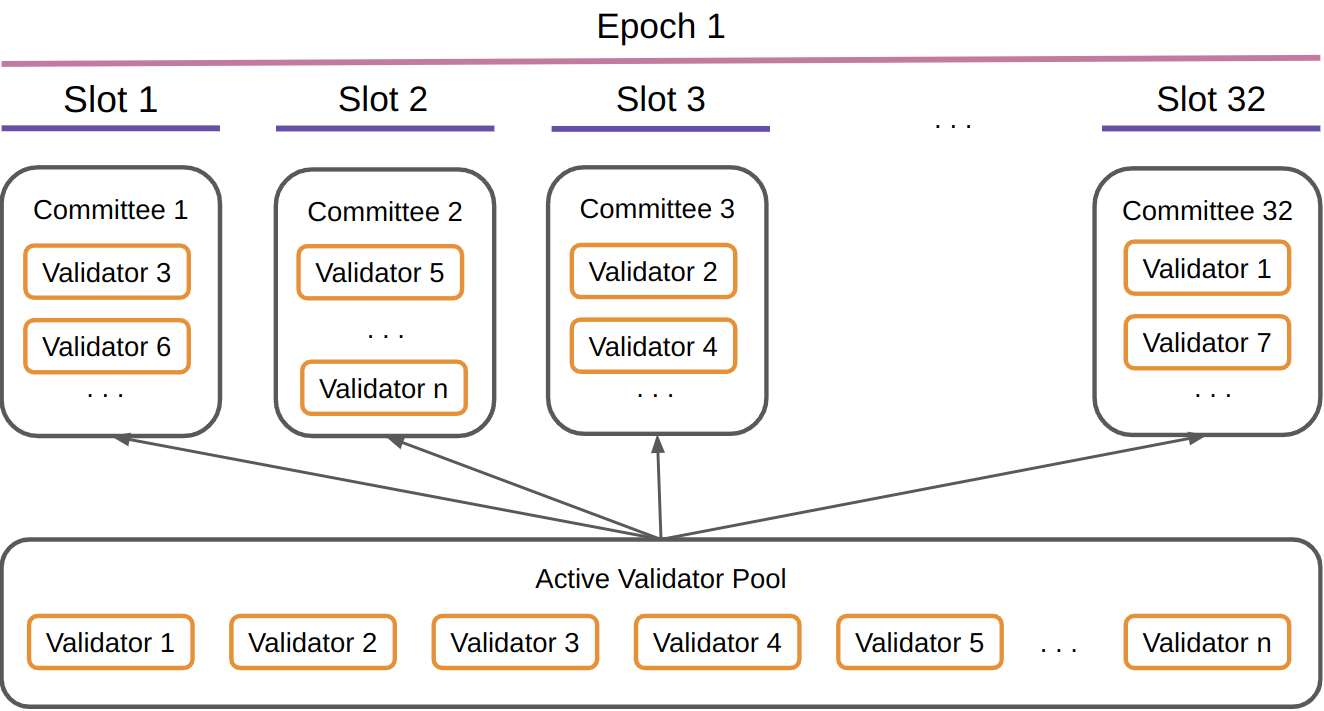}
        \caption{Assignation of active validators into attestation committees.}
        \label{fig:ethereum_committees}
    \endminipage\hfill
    \minipage{0.45\textwidth}%
        \includegraphics[width=\linewidth]{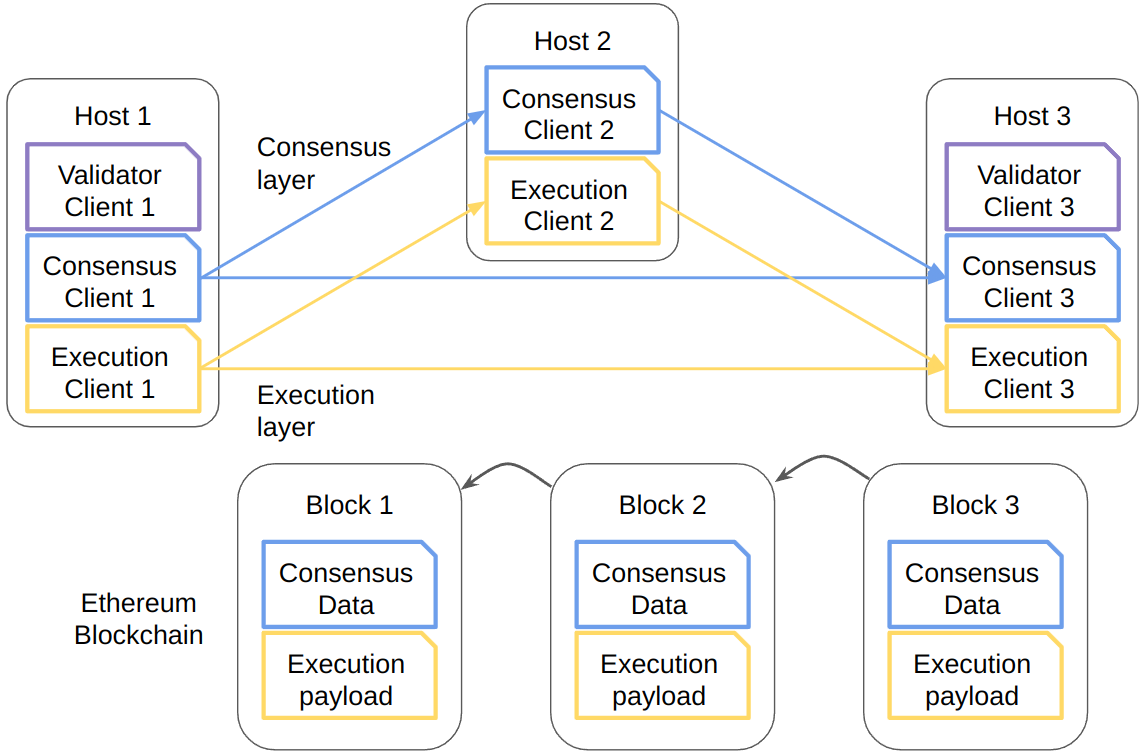}
        \caption{Subdivision in layers of Ethereum after the merge.}
        \label{fig:layers_in_ethereum}
    \endminipage
\end{figure*}

In the beacon chain, only active validators can participate in block proposals and attestation committees, assuming the duty and responsibility to participate in the consensus over each proposed block.
Based on the previous state of the chain and the balances, a randomly chosen validator has a time window of 12 seconds to propose a new block. 
This time window, or slot, also represents the time range other committee participants had to perform their attestations (visually represented in Figure \ref{fig:ethereum_committees}). If a validator does not perform its duties, it gets exposed to economic penalties, while honesty and participation gets rewarded.
The current PoS model requires the participation of $2/3$ of the active validators to consider previous blocks finalized (immutable, as they achieve enough support from the validators). In fact, the network organizes the slots in epochs, where each epoch includes 32 slots. 

At the first stage of the transition, the only link between both blockchains was the activation smart contract. But later on, with the Merge on September 15th, 2023, both chains merged into a single one (see Figure \ref{fig:eth-transition}).
However, as Figure \ref{fig:layers_in_ethereum} shows, the blockchain has two distinct layers: i) the execution one and ii) the consensus one, remaining with one blockchain, one block on each slot, and two layers underneath.
The Execution Layer (EL) works as the previous Ethereum network, where users broadcast transactions interacting with the EVM.
On the other hand, the Consensus Layer (CL) keeps track of the duties, balances, state, and performance of each validator in the beacon chain.

As each layer has specific requirements, each layer has specific software or clients to participate in. However, each layer’s clients can't work alone without a direct pair from the other layer, creating a tandem. Even though the execution layer node could receive the transactions over the network, it won't certainly know which is the last finalized slot that defines the balance at a given Ethereum address. In the same way, if the consensus client doesn't have a connection with the execution layer client, it won't be able to validate the transactions from any proposed block or even fill up blocks with valid transactions if a hosted active validator becomes a block proposer.
On top of that, if a user or the entity wants to contribute to the chain's consensus, they will have to plug a third client on top of the beacon node (consensus layer client), a validator client. This third client will be responsible for signing the validator duties for each validator it hosts, as it can run thousands of validators under the same beacon node. The idea behind running a third software on top of the beacon node is to preserve the anonymity of validators in the peer-to-peer network, which makes targeted attacks harder as there is no direct link between which beacon node hosts which validators.

In the original track of the beacon chain \cite{og-sharding}, there was a second underlined solution to tackle one of the most common limitations of blockchain protocols: the protocol's scalability. 
The original idea of increasing the scalability of Ethereum was the addition of blockchain shards on the consensus layer, where the whole list of active validators could split across the available shards and limit their activity to the same one. This sharding schema was postponed due to the large complexity of the model. However, thanks to the current stability of the network after the merge, new sharding alternatives have been proposed, i.e., data sharding \cite{sel2018towards} \cite{hall2023foundations}.

The approach to upgrade Ethereum's consensus mechanism references the previous PoW consensus mechanism. PoW is known for being a resource-intensive consensus mechanism \cite{denisova2019blockchain}, where users need to invest significant amounts of money on hardware to increase their chances of being a block proposer, thus being more profitable.
However, this also means that, as the hardware improves the mining capabilities over time \cite{mahony2019systematic}, the more obsolete the previous hardware becomes. This highly incentivizes the acquisition of newer mining devices, indirectly increasing the network's overall energy consumption \cite{li2019energy} over time.

Ethereum's transition to PoS has been highly motivated by this drawback. Although you still need to invest money to be profitable (you need $32$ Ether to activate a validator), the hardware requirements to run multiple validators barely change from the node's perspective. One single beacon node can host up to thousands of validators.
The motivation of the community and the core developers is clear: try to reduce the requirements to run a node so that it becomes more accessible, improving the sustainability aspect and the resilience of the network, as having a full Ethereum node at home became less restrictive (from an infrastructure point of view).

This paper thoroughly analyzes the hardware requirements the main beacon chain clients need over the different stages of their lifetime and over multiple hardware combinations. We empirically demonstrate and quantify the minimum resources needed to run a full Ethereum node after the merge, and we extensively associate the relation of each event in the network with the specific hardware it requires. 
By doing this, we present i) the impact of the different approaches each client developer team made into the user’s hardware, ii) the importance of these measurements to spot bugs and misbehaving clients under specific network states, and iii) the possibility of analyzing the healthiness of the chain in past periods of time only by looking at the synchronization of the chain at that particular time range. 
We would like to mention that this work has been carried out in close collaboration with the Ethereum Foundation and the client core developer teams. Some of these teams have benefited from this collaboration by identifying bottlenecks and implementing some of the improvements raised by this study. The contributions of this study have also been considered interesting by the Ethereum Foundation, which has asked to perform it continuously, offering a real-time dashboard that can help future users understand which client must be the best fit for their needs.

The remainder of this paper is organized as follows. Section~\ref{sec:related-work} discusses related work.
Section~\ref{sec:methodology} explains the methodology used for the evaluation.
In Section~\ref{sec:evaluation}, we introduce and analyze the results obtained by our study.
Section \ref{sec:discussion} discusses the main findings of the paper.
Finally, Section~\ref{sec:conclusion} concludes this work and presents some future directions.

\section{Related Work}
\label{sec:related-work}

PoW’s security comes with the aggregation of more hashing power to the network since more people honestly participating means more resources an attacker has to invest to achieve 51\% of the mining power. However, this is a double-edged sword, as the hardware requirements and the energy consumption increase as more users participate in the PoW consensus mining.

This was clear first in Bitcoin, and later on in Ethereum, where the total hashing power of the network has been constantly increasing over time \cite{taylor2017evolution} \cite{wang2015exploring}. There has been a lot of research to optimize and speed up the process of mining\cite{shen2014look}, first with the usage of GPU accelerating techniques, and reaching, in some cases like Bitcoin, the development of custom ASICs \cite{mahony2019systematic}. 

However, as we have already pointed out before, this comes with a significant drawback, as the scalability remains the same while the network requires more resources. 
Ethereum’s PoW network’s security and vulnerabilities have been highly tested in the past \cite{chen2020survey}. But the transition to PoS adds more complex logic to the already existing protocols, as this mechanism presents a few critical points that could potentially be exploited in the future.

In the first place, blockchains relying on proof of stake consensus mechanisms highly depend on the network latency \cite{Moindrot2017ProofOS}, and Ethereum is not an exception \cite{cortes2023unveiling}. Whiling to have hundreds of thousands of messages distributed in a time window of $12$ seconds puts a lot of pressure on its peer-to-peer networking layer \cite{kim2018measuring}.

Finally, previous works have shown the importance and effectiveness of monitoring the hardware resource utilization to identify memory leaks, bugs, and edgy case scenarios that can become a bottleneck for the correct performance \cite{inagaki2019profile}. 
Of course, there is a direct link between the hardware utilization and the overall energy consumption of the machine. Larger hardware needs come at a cost: more energy-hungry hardware. Thus, monitoring these requirements can help understand the overall energy consumption of the software and the network \cite{noureddine2015monitoring}. This is the case of Ethereum, coming from a former PoW scenario, where monitoring the resources of PoS clients can showcase the energy reduction that PoS means to the total carbon fingerprint of the network.   

As the resilience of the network now relies upon the online and honest behavior of the validators \cite{cortes2023autopsy}, the software that runs the PoS logic and interacts with the rest of the network becomes a more central piece in the equation.
While the execution clients were analyzed in the past~\cite{rouhani2017performance}, in this paper, we study the resource utilization of consensus clients. In particular, we aim to fill the existing lack of a hardware resource analysis of the first transition of consensus mechanism in a live network. In the paper, we present the evolution of the hardware requirements from the available software to participate in Ethereum's PoS. We showcase the vast reduction of the requirements, deny some myths around the minimal requirements to run an Ethereum node and present the possibility of predicting what is going on in the chain just by checking the allocated resources from the clients.

\section{Methodology}
\label{sec:methodology}

To study differences in the behavior of the Ethereum CL clients, we have monitored the hardware resource utilization of the six main consensus clients, Teku \cite{teku}, Nimbus \cite{nimbus}, Lighthouse \cite{lighthouse}, Prysm \cite{prysm}, Lodestar \cite{lodestar}, and Grandine \cite{grandine}, in different machines with the most common hardware combinations. 
Among the studies we have performed, we can distinguish two main phases to monitor the run-time of a client: the synchronization phase and the chain-head following phase. 
In the first phase, the clients aim to reach the latest state of the chain. Generally, by syncing and verifying each of the blocks in the chain from the Genesis block, clients have to re-compose the chain by asking peers in the network to share the blocks with them. To finally reach the point where they are already up with the head. This process might be more stressful for the machine, as most of the time, the client has to download, process, and verify as many blocks as possible in the shortest possible time.  
In the second phase, clients are less stressed, as the number of blocks to process happens once every 12 seconds.  

By monitoring the hardware resources of consensus clients in different types of machines, this paper aims to track and understand the client's needs, limitations, behaviors, and, thus, performance while participating in the Ethereum network. To do so, the metrics that have been monitored are:
\begin{itemize}
    \item CPU
    \item Memory Usage
    \item Disk Usage
    \item Network outgoing traffic
    \item Network incoming traffic
    \item Syncing time
    \item Peer connections
\end{itemize} 

\subsection{Different users, different hardware}
\label{subsec:control-hardware}
To achieve a reasonable comparison of the different clients and, even more importantly, to understand and highlight the minimum requirements for running an Ethereum beacon node, we have selected three different sets of hardware combinations that, from our perspective, summarize the different users that would run a beacon node in Ethereum: \emph{enthusiasts}, \emph{solo stakers}, \emph{staking companies} or \emph{node operators}. 

\begin{itemize}
    \item The first category, \emph{enthusiasts}, are individuals who follow the development of the Ethereum protocol and want to actively support it by running a node in low-power or energy-efficient devices. They don't necessarily need to run a validator but want to find a productive task for their ``dusty" \emph{Raspberry Pi}.
    \item Solo stakers, on the other hand, are passionate individuals who want to contribute to the chain's consensus and run one or a few validators at their own place. They generally have a dedicated personal machine such as an \emph{Intel NUC} or a built PC with the standard components for the personal PC market.
    \item The final target user is the most professional-oriented one, involving users like \emph{staking entities} and \emph{node operators}. These network participants generally participate in the network with a lucrative interest. They generally run hundreds to thousands of validators. Thus, they don't stay short on hardware resources to run them.
\end{itemize}

\begin{table}
    \centering
    \begin{tabular}{ccccc}
        \hline
        Name & CPU & Memory & Storage & Network \\
        \hline  
        Raspberry Pi 4b  & 4c   & 8GB      & 256GB     & 100Mbit/s     \\
        Default node     & 4c.  & 15GB     & 100GB SSD & 250Mbit/s     \\
        Fat node         & 32c. & 120GB    & 400GB SSD & 10.000Mbit/s  \\
        \hline
    \end{tabular}
    \caption{Hardware configuration of the control machines.}
    \label{tab:hardware-configuration}
\end{table}

The hardware resources chosen to test the clients are summarized in table \ref{tab:hardware-configuration}.
Each of the six main clients has run independently (alone on a dedicated machine) on the three available platforms. Of course, the testing times highly depend on the synchronization speed of each client and platform, and each of the three platforms has been monitored on separate dates between 2022 and 2023. Each respective performance phase and platform has its independent subsection in section \ref{sec:evaluation}, where the analysis of the results is extended.  

\subsection{Node exporter}
\label{subsec:node-exporter}

Each control machine we have used to perform the study has been entirely monitored by the software tool \emph{Node Exporter} \cite{node-exporter}, giving us access to the whole list of metrics mentioned in the previous paragraphs. Combined with a central \emph{Prometheus} \cite{prometheus} time-based database service that scrapes each of the exported endpoints every 15 seconds, it can accurately provide the resources in use for an entire machine.

As calibration for the software tool, we have benchmarked the results obtained from the node exporter with the ones taken from a \emph{Python script} that reads the same metrics in run-time from the machine every second. The successful comparison between the values of each gathering tool demonstrated that the node exporter showed a negligible overhead to the measurements while providing the confidence of using a reliable, unbiased, and well-tested tool to monitor the resources of each machine.

\section{Evaluation}
\label{sec:evaluation}

To extend into a larger and more detailed analysis of how the Ethereum PoS transition has impacted the hardware requirements to participate in the network and consensus, we have accumulated almost 30,000 CPU hours (3.4 CPU years) across Ethereum's CL clients in different hardware and configurations. The study includes data from over a year, where we have collected over 735 million data points, from which we extracted almost 150 million data points to plot about a thousand different figures showing how the other CL clients perform. 

The following section discusses our findings, reviewing the most critical points that we identified in our journey into Ethereum's PoS transition. We have summarized or compressed the graphs and plots to the minimum for time and space reasons. However, following the transparency and open-source standards of the Ethereum community, all the figures are accessible on the following GitHub repository ~\cite{migalabs-resource-website}.

\begin{figure*}[]
    \minipage{0.47\textwidth}
        \includegraphics[width=\linewidth]{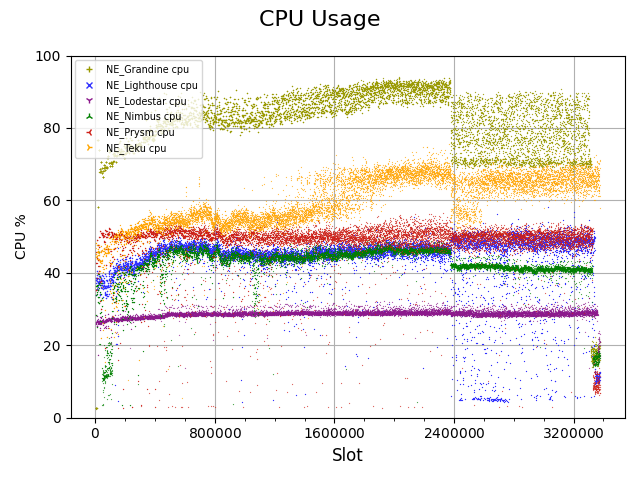}
        \caption{CPU utilization while syncing the chain from a default node. Comparison across clients.}
        \label{fig:default-cpu}
    \endminipage\hfill
    \minipage{0.47\textwidth}%
        \includegraphics[width=\linewidth]{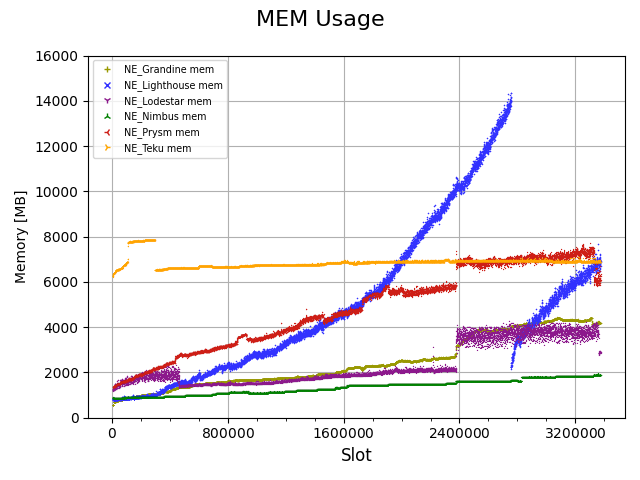}
        \caption{Memory utilization while syncing the chain from a default node. Comparison across clients.}
        \label{fig:default-mem}
    \endminipage
\end{figure*}

\subsection{Synchronization process}
\label{subsec:sync}

\begin{table}[]
    \centering
    \begin{tabular}{ccc}
        \hline
        Client & Version \\
        \hline  
        Prysm & 2.0.6 \\
        Lighthouse & 2.1.4 \\
        Teku & 22.3.2 \\
        Nimbus & 1.6.0 \\
        Lodestar & 0.34.0 \\
        Grandine & 0.2.0 \\
        \hline
    \end{tabular}
    \caption{Table with the versions used during the \emph{default node} study.}
    \label{tab:default-versions}
\end{table}

The full synchronization of the chain is the event of downloading and verifying the historical blockchain from the genesis block to the head of the same one. It is a process that only gets done at the beginning of the client's lifetime, and that could be repeated on rare occasions, such as when this client is changed to a different one. Nevertheless, it is still essential for the network to ensure the existence of full nodes in the network with all the historical data downloaded. Thus, it is vital for the community to monitor the chain synchronization from Genesis, as it can help users better estimate the preparation time of each node before they can host any validator. 

Full nodes ensure the blockchain is composable and verifiable at any given time. Therefore, it is essential to ensure that the process is not slow, long, and tedious if users want to sync the entire chain from scratch. Especially when the downtime of an already activated validator might depend on the downtime of the Beacon Node. To measure the different synchronization techniques proposed by the community, we have monitored this process for the main clients in the Ethereum CL ecosystem.

The following evaluation paragraphs refer to the gathered data when syncing the chain from genesis on the six clients on \emph{default mode}. The data was collected until the 18th of March, 2022, with the following client versions shown in Table \ref{tab:default-versions}.
It is essential to mention that by the time we made this measurement, it wasn't necessary to pair each CL client with an EL client to keep the head of the chain correctly. Thus, all the metrics displayed in this section will refer only and exclusively to the resources used by the CL clients.

\subsubsection{CPU utilization} 
\begin{figure*}[!htb]
    \minipage{0.45\textwidth}
        \includegraphics[width=\linewidth]{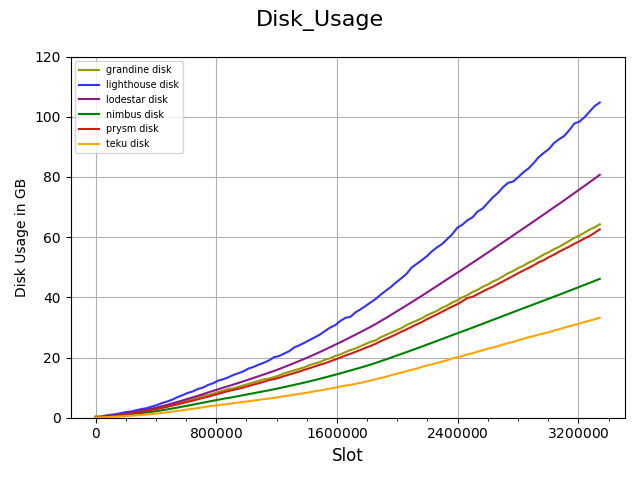}
        \caption{Disk usage while syncing the chain from a default node.}
        \label{fig:default-disk}
    \endminipage\hfill
    \minipage{0.45\textwidth}%
        \includegraphics[width=\linewidth]{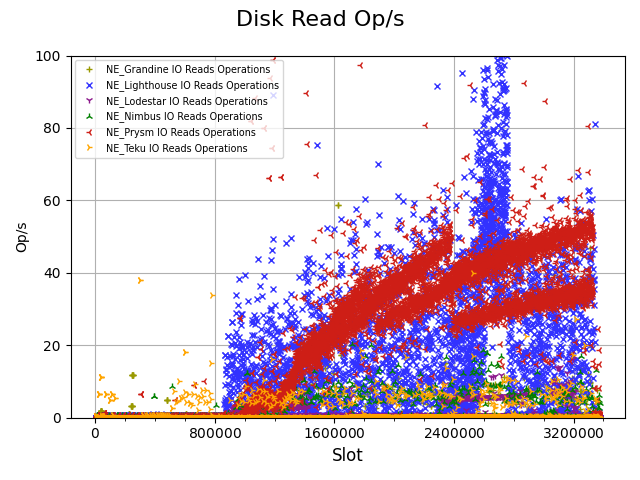}
        \caption{Disk read operations/s while syncing the chain from a default node.}
        \label{fig:default-disk-reads}
    \endminipage
\end{figure*}

The implementations of Ethereum's CL specifications are written in different programming languages, which can ultimately determine critical parameters such as the level of concurrency the program can achieve or the optimization level for the validation and underlaying processes. Figure \ref{fig:default-cpu} shows the CPU utilization degree achieved by the CL clients using the chain slots as a reference; this helps to have a fair comparison between clients, as some finished sooner than others. 
In the figure, we can appreciate different CPU profiles, where Grandine follows a different recognizable pattern, using around the $80\%$ of the CPU while syncing. Teku follows the lead on CPU utilization with a slightly increasing ratio reaching the $60\%$ of the CPU. The figure shows similar CPU profiles for Prysm, Lighthouse, and Nimbus, leaving Lodestar as the client that requires less CPU to synchronize the chain.

There is a crucial point to highlight here; although we could associate a high CPU utilization with a drawback to a CL client (we come from a premise where reaching consensus in PoS requires very few resources), it isn't necessarily bad. Not at least if the CPU cycles are optimized to sync the chain faster. Remember that the sooner we can sync the chain from Genesis, the lower the downtime users could experience on their validators. Thus, we support the idea of squashing the available resources during the synchronization phase if it reduces the duration of the same one.

\subsubsection{Memory utilization} 
In terms of memory, all the clients expose different behaviors. Figure \ref{fig:default-mem} shows the memory allocation patterns from the CL clients. The figure shows that most of them allocate more memory throughout the synchronization process. This is an expected pattern, as the beacon state \footnote{The beacon state represents the state of the chain at a given slot. It includes the status, balance, and information of each validator.} grows as time passes, and more validators tend to join the chain. Among all the recorded patterns, it is clear that Lighthouse has an unusual, constantly increasing one. Reaching even the limits of the machine, what we can identify as a possible memory leak, leading to a crash and a forced restart of the machine. Of course, the incident was shared and reported with the Sigma Prime team \cite{sigp} (developer team of Lighthouse), and it is known to be fixed in the following version $v2.2.0$. 

On the other hand, it wasn't the only client that reported some difficulties when setting it up. Teku and its associated JVM are somehow tedious to configure. The JVM requires a minimum amount of memory to work, experiencing sudden crashes if not enough memory is provided. However, we could achieve a steady performance by assigning $6GB$ of memory to the JVM. As represented in the figure, Teku keeps a reasonably constant Memory utilization, never exceeding $7GB$ of memory.
The memory profiles get more stable with the rest of the clients, where only Prysm reaches the same level as Teku, and Lodestar, Grandine, and Nimbus keep a lower profile at a lower limit of $4GB$. It is worth mentioning that Prysm is written in \emph{Go}, which tends to allocate and keep the memory for the processes until the OS asks it back. Thus, this makes the comparison a bit unfair for Prysm, as part of the measurement might belong to unused but still allocated memory by \emph{Go}. In these lower memory profiled clients, Nimbus has the lowest memory consumption, with a steadily increasing profile that keeps under $2GB$ of memory allocation and shows an incredible memory optimization for such an intense process.

Similarly to the CPU utilization profile, we can identify where most clients start to behave differently. Around slot number 2.4 million, this point refers to the transition to Altair's hard fork in the Beacon Chain. From Altair, the CL clients need to track sync committees and other significant changes in slashing conditions, which explains the difference in resource consumption. The overall block size also increased, making it heavier to download and slower to process while keeping more bytes in memory.   

\begin{figure*}[!htb]
    \minipage{0.47\textwidth}%
        \includegraphics[width=\linewidth]{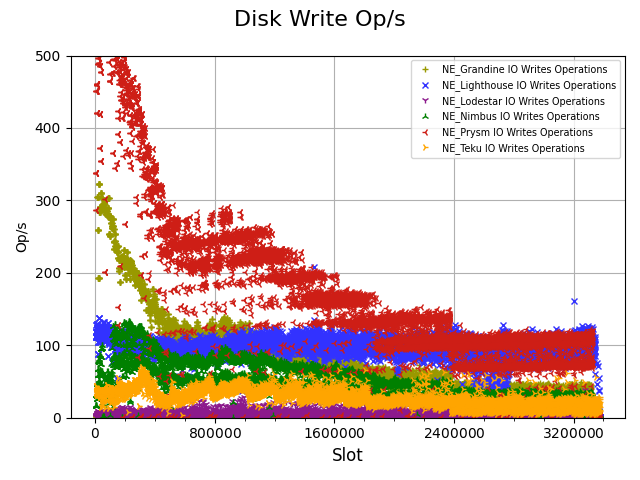}
        \caption{Disk write operations/s while syncing the chain from a default node.}
        \label{fig:default-disk-writes}
    \endminipage\hfill
    \minipage{0.47\textwidth}
        \includegraphics[width=\linewidth]{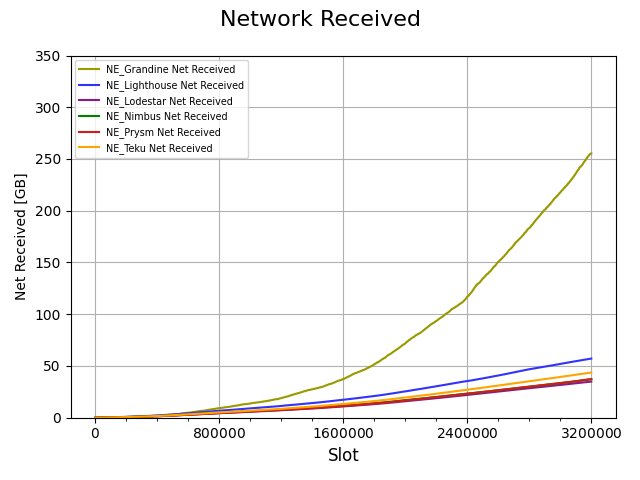}
        \caption{Network inwards utilization while syncing the chain from a default node.}
        \label{fig:default-netin}
    \endminipage
\end{figure*}

\subsubsection{Disk utilization}

The differences are pretty impressive when comparing the distinct size it took to keep the entire chain by the clients. Figure \ref{fig:default-disk} shows the average disk usage using the slots as a reference. Although all the clients have run on their default configuration, the figure shows that there is an important difference in how much storage the different clients require; for instance, Lighthouse takes over three times the storage of Teku, which is the CL client that requires the least storage followed by Nimbus. There is a catch here: although all the clients are in charge of keeping the whole set of raw blocks on disk (raw blocks databases barely change in size across clients), there are differences between the frequency of maintaining beacon state checkpoints. 
We have already mentioned that the beacon state represents the chain snapshot at a given slot. It changes over time as more blocks are added to the chain. Thus, there is no need to keep it in disk with a super low frequency, as its size generally is much larger than a single block. For this reason, beacon nodes keep track of beacon state checkpoints, and when they need to access a specific state from the past, they can load the closest state they have stored, applying the subsequent list of blocks until they can regenerate the desired state.   

All this said, we can deduce from Figure \ref{fig:default-disk} that the default checkpointing in Lighthouse has a higher frequency, keeping in disk more checkpoints for the same range of slots, which ultimately helps access faster any state in the past. It is important to note here that this parameter is adjustable in all the clients, and it might be important to tune up based on the necessities and resources of each user.

The disk utilization has some other peculiarities, though. 
With a wider focus on disk write operations in Figure \ref{fig:default-disk-writes}, most clients, and in particular Prysm, have a higher number of disk write operations per second at the beginning of the syncing process, and this decreases gradually until it plateaus after Altair.
This is explained by the fact that at the beginning of the Beacon chain, the number of validators (21.063) was significantly lower than 3.2 million slots later (295.972), making the whole slot processing of the blocks much faster.
Both figures \ref{fig:default-disk-writes} and \ref{fig:default-disk-reads} are inversely correlated; the increase in disk reads is relatively low in the first half and accelerates in the second half, as the number of validators and attestations in the network increases.
The measurements in Figure \ref{fig:default-disk-reads} show that disk read operations increase dramatically for most CL clients, except for Teku. This is not an issue because it does not affect the performance. But it happens virtually simultaneously for most clients around slot $900K$, being Prysm client to more clearly expose this behavior.

We attribute this pattern to some falling caching techniques that might force the client to read the needed information out of the internal client cache, which produces more read operations.  

\begin{figure*}[!htb]
    \minipage{0.47\textwidth}%
        \includegraphics[width=\linewidth]{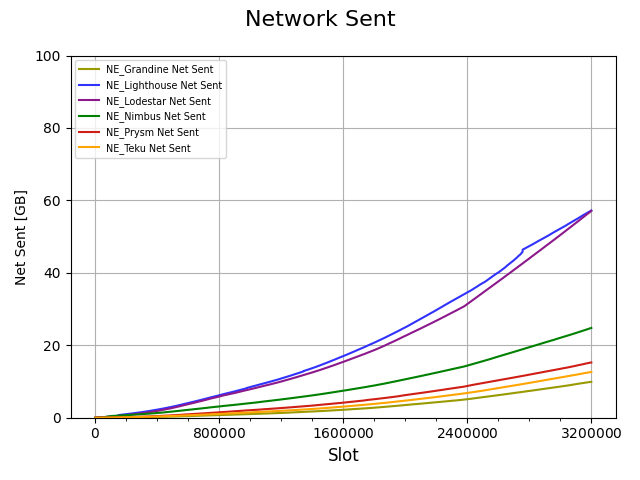}
        \caption{Network outwards utilization while syncing the chain from a default node.}
        \label{fig:default-netout}
    \endminipage\hfill
    \minipage{0.47\textwidth}
        \includegraphics[width=\linewidth]{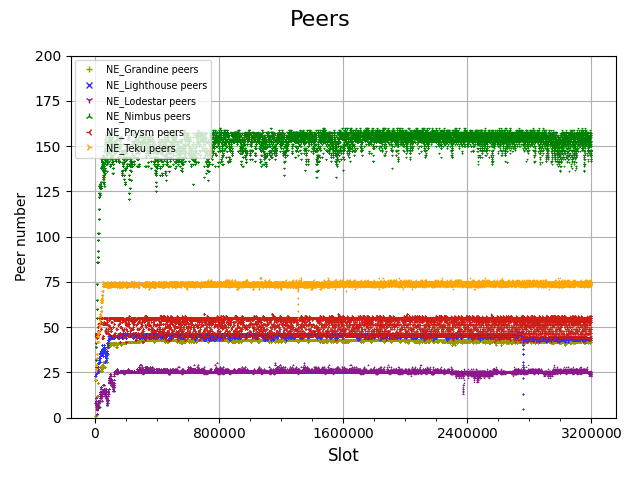}
        \caption{Number of connections while syncing the chain from a default node.}
        \label{fig:def-peers}
    \endminipage
\end{figure*}

\subsubsection{Network bandwidth}

Network connectivity in a distributed network is a parameter highly associated with the number of concurrent connections the node keeps with others in the network. Despite being in the synchronization phase, Ethereum is not different from other networks in that aspect. Figures \ref{fig:default-netin} and \ref{fig:default-netout} show the aggregated received and sent \emph{GB} throughout the synchronization of the chain.

Figure \ref{fig:default-netin} shows that most of the clients follow a similar ratio of total downloaded GBs to sync the chain around the total of 50GB. Grandine, on the other hand, outstands its competitors, requiring five times more bandwidth to perform the same task. It seems that handling more concurrent downloads to sync the chain faster comes to a not-that-optimized deduplication of downloading multiple times the same blocks.

Regarding sent GBs, Figure \ref{fig:default-netout} shows that only Lodestar and Lighthouse are above the rest of the clients with around five times more output bandwidth than the ``quieter" clients.

As mentioned, the bandwidth usage is highly attached to the number of simultaneous peers each client has. Figure \ref{fig:def-peers} shows the number of peers each client got during the syncing process. The figure shows that by default, Nimbus looks for stable $150$ peers from where to fetch blocks. Followed by Teku with $75$ and Prysm with $50$, and the pair of Lighthouse and Grandine, which share a peer target of $45$. To finish, Lodestar has shown the lowest ratio of clients with a stable target of $25$ concurrent peers.

\subsubsection{Performance}

We have presented many insights about how each client performed over this first step of connecting to Ethereum's network and synchronizing the chain. We have seen many different design choices to perform this same task, like choosing to keep more data in the cache to reduce the number of interactions with the disk (the slowest task of a computer), increasing the number of concurrent processing tasks to minimize the total time, or optimizing the process to make it as light and fast as possible.

\begin{figure*}[!htb]
    \minipage{0.47\textwidth}%
        \includegraphics[width=\linewidth]{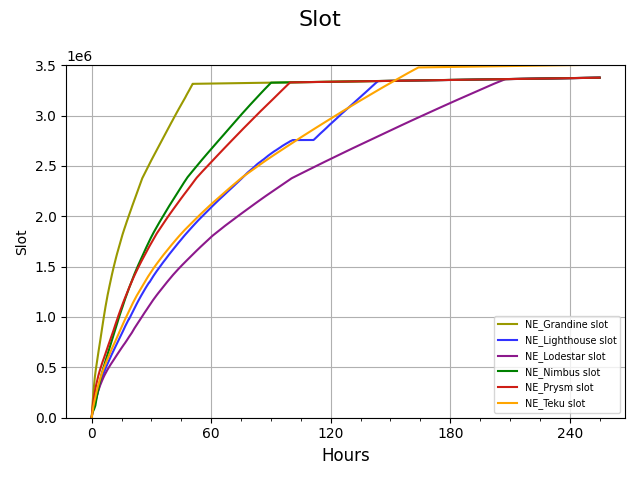}
        \caption{Chain synchronization speed of the blockchain from a default configured node.}
        \label{fig:def-sync-slots}
    \endminipage\hfill
    \minipage{0.47\textwidth}
        \includegraphics[width=\linewidth]{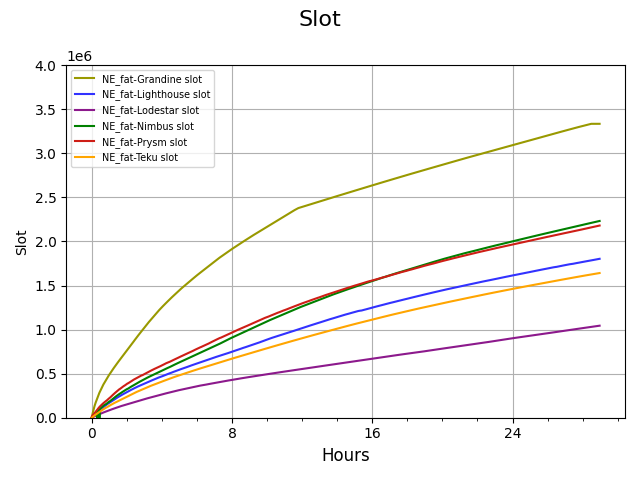}
        \caption{Chain synchronization speed for the different clients on a fat node.}
        \label{fig:fat-sync}
    \endminipage
\end{figure*}

Figure \ref{fig:def-sync-slots} shows each client's total duration of the synchronization process. The figure shows that concurrency has clear benefits in speeding up the process. Grandine was the fastest client to catch up with the head of the chain in almost $50$ hours. The fastest clients are closely followed by Nimbus and Prysm, with around $90$ hours, displaying that it is possible to achieve great timings if optimization, concurrent downloads from multiple clients, and a mild level of CPU usage are achievable. Lighthouse and Teku have also achieved similar results with around $135$ to $150$ hours-long process. It is not a fair comparison for Lighthouse, as it crushed for an over-allocation of memory and had to be restarted. However, the figure clearly shows that it might be the most trustless client of the set, investing in keeping more states on disk more often in case a chain reorganization happens. Teku and Lodestar seem to be the least optimized clients catching up with the chain's head. Despite the highest memory allocation, the second highest CPU utilization, and the second highest number of peers from Teku, it still got in fifth place. In the last place, the measurements show that Lodestar has one of the lowest profiles of the clients, perhaps aiming for a light client that could run in the background.

\subsection{The bigger, the better?}
\label{subsubsec:fat-nodes}

As anyone could expect, having a more capable machine has its benefits. The majority of the tested clients take advantage of parallelization techniques to speed up the entire synchronization process. However, some implement concurrency techniques better than others, taking a huge advantage as they can sync up the historical chain sooner. With that many parameters involved in the performance of a client, increasing each of them individually can vary the benefits differently:
\begin{itemize}
    \item A faster CPU with more cores allows the client to process attestations at a higher rate and, therefore, more blocks simultaneously. However, there is a point that having many CPUs is not useful since many cores may stay idle if there aren't that many events to process and validate. 
    \item A bigger memory available while syncing increases the number of items the client can keep without performing slow read and write operations to disk.
    \item A higher network bandwidth can allow us to keep more simultaneous connections with peers in the network, enabling the concurrent download of more blocks from the chain.
    \item A faster disk can reduce the bottleneck originated by writing such a long chain as the Beacon Chain.
\end{itemize}

\subsubsection{Sync Speed}

Our measurements show that machines with superior hardware resources can synchronize the chain faster. However, with the apparition of checkpoint syncing, the whole operation of syncing the chain from Genesis can be reduced to almost zero times. Some CL clients offer a back-filling sub-process to fetch the historical chain backward from the given checkpoint. However, this is not the case for all of them, and if users want to have faster access to on-chain data, known as ``archival node", they might still be forced to perform a full synchronization from Genesis. 
This ``archival" mode increases the frequency at which the chain checkpoints are stored in the database, making all the RPC queries generally faster as it takes less time to recompose intermediary states. Making it ideal for those users who want to have faster access to chain data.  
The whole concept of checkpoint syncing relies on the node's single need to access the last beacon's finalized state to operate. This process allows a node to fetch the last finalized Beacon State and Signed Beacon Block from a trusted node, allowing it to process and validate incoming new blocks after the process has been accomplished\footnote{Although the beacon node or the CL client can process new incoming blocks after a checkpoint sync, its performance is still subjected to the synchronization of the paired EL client. EL clients, when writing this paper, can not sync from a checkpoint state. Thus, the time needed to have an operative EL + CL client relies on how fast the EL can catch up with the head of the chain.}. Nevertheless, although checkpoint-syncing is the recommended operation to catch up with the chain's head, not all users can access an already synced beacon node from where to fetch the last finalized checkpoint, making the full synchronization speed still important.  


Figure \ref{fig:fat-sync} shows that clients achieve a better synchronization speed with more capable hardware. On average, our measurements show a speed improvement of $x1.352$, with Lodestar showing a disappointing slower performance of $x0.90$ and Grandine with a remarkable $x1.785$ times faster synchronization. All the measured slot processing ratios per second are displayed in Figure \ref{fig:fat-sync-slots}, where we can appreciate the average performance of Teku, Nimbus, Prysm, and Lighthouse, and the outlying one from Grandine and Lodestar.

\subsubsection{Disk utilization}
\begin{figure*}[!htb]
    \minipage{0.47\textwidth}%
        \includegraphics[width=\linewidth]{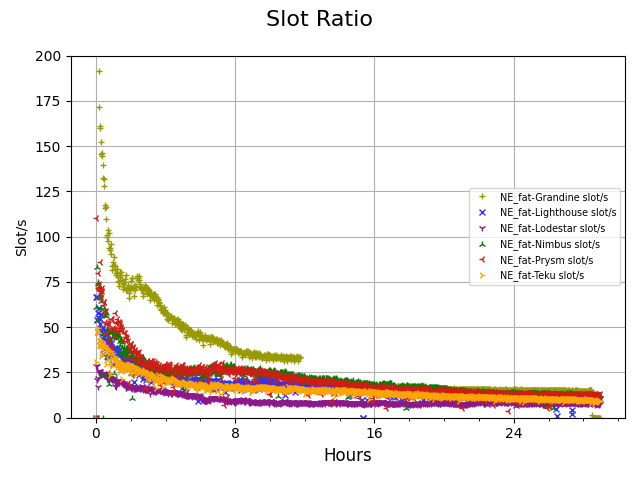}
        \caption{Slot synchronization ratio per second for the different clients.}
        \label{fig:fat-sync-slots}
    \endminipage\hfill
    \minipage{0.47\textwidth}
        \includegraphics[width=\linewidth]{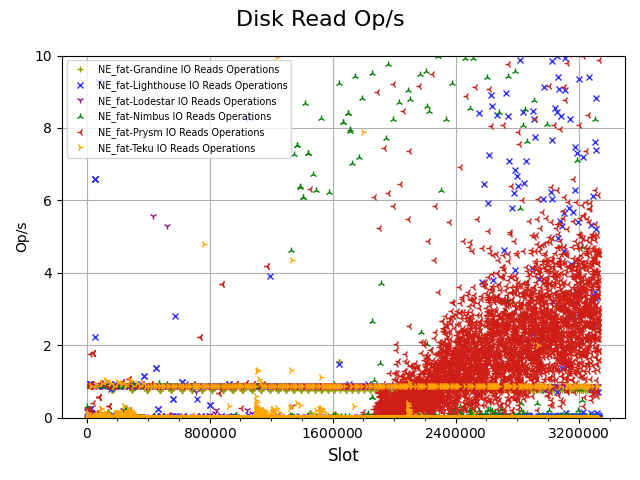}
        \caption{Disk read operations by the clients while running in a ``Fat" node.}
        \label{fig:fat-disk-reads}
    \endminipage
\end{figure*}

Of course, having more and faster processing power implies more disk reading and writing operations (if no disk bottleneck is hit). In the previous subsection \ref{subsec:regular-performance}, we appreciated how around slot $900.000$, the disk reads spike until the end of the process. Controlling the same metric on the fat nodes, Figure \ref{fig:fat-disk-reads} shows almost the same behavior, but this time, the pattern started much later, around slot $2.000.000$. The fact that the phenomenon occurs later on nodes with more memory makes us believe some memory caching process is originating or preventing these disk read operations. For this reason, when the client reaches its buffering limit, it starts generating a significant amount of disk reads.

On the other hand, disk write operations remain on the same pattern as Figure \ref{fig:fat-disk-writes} represents, increasing its offset with the faster synchronization speed. This showcases that although we need a proper or fast enough disk to handle an Ethereum CL node, unless we want to set up an archival node, there is not much impact on disk usage originated by increasing the rest of the hardware components.

\subsection{How small could we go?}
\label{subsec:raspi-sync}
Despite the complexity increase that shifting from PoW to PoS means, the protocol has bragged about requiring less computational power to reach consensus than its PoW predecessor version. Going even to the limit of stating that a Beacon client could run on a Raspberry Pi. We have run all the different clients from scratch in Raspberry Pis on the versions displayed in Table \ref{tab:default-versions}, testing if running a production node in such a low-powered device is possible. Anticipating the slowest performance of the more limited resources of the Raspberry Pis, the syncing measurement of the chain for the low-powered devices was performed on the \emph{Kiln} testnet \cite{kiln-testnet} on the same range of dates (18th of March, 2022). 

\subsubsection{CPU utilization} 

\begin{figure*}[!htb]
    \minipage{0.47\textwidth}%
        \includegraphics[width=\linewidth]{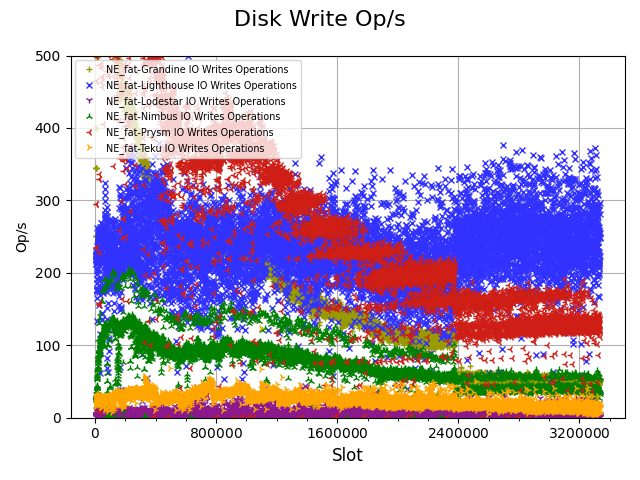}
        \caption{Disk write operations by the clients while running in a ``Fat" node.}
        \label{fig:fat-disk-writes}
    \endminipage\hfill
    \minipage{0.47\textwidth}
        \includegraphics[width=\linewidth]{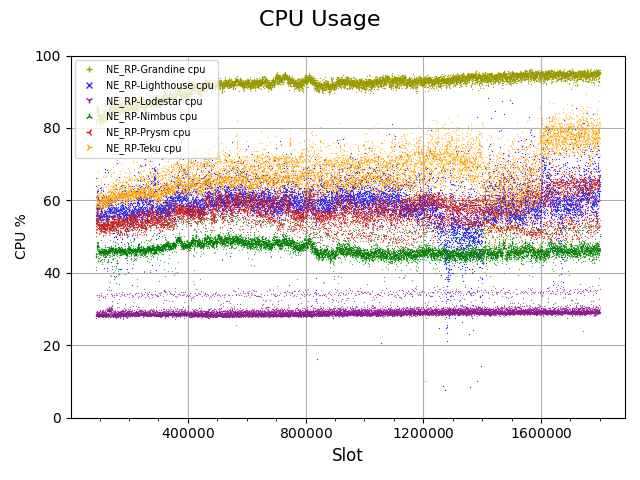}
        \caption{CPU utilization while syncing the chain in a Raspberry Pi 4.}
        \label{fig:raspi-cpu}
    \endminipage
\end{figure*}

Figure \ref{fig:raspi-cpu} shows the CPU utilization degree achieved by the CL clients, which remains relatively similar to the one measured on the default node with a slight overhead as each core is ``slower" than its \emph{default} version.
The figure shows that Grandine increases CPU utilization from $80\%$ to a constant $90\%$ CPU usage while syncing. Teku follows the lead, increasing its CPU utilization up to the $80\%$ as the blocks keep getting bigger due to the aggregation of more validators. At the same time, the rest of the clients also seem to register an increase of the $10\%$ of the usage.

\subsubsection{Memory utilization} 

\begin{figure*}[!htb]
    \minipage{0.47\textwidth}%
        \includegraphics[width=\linewidth]{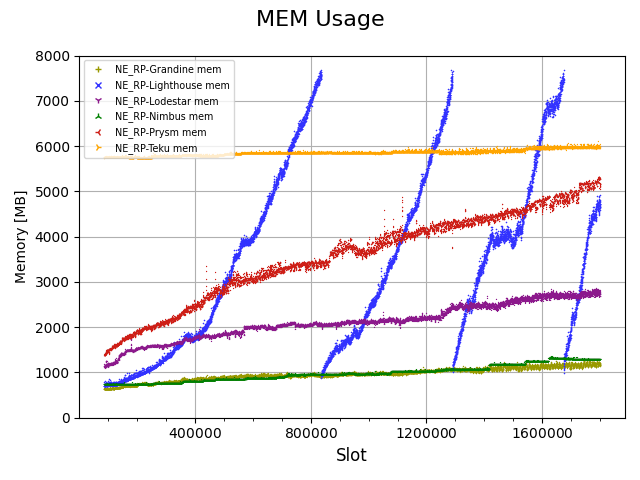}
        \caption{Memory utilization while syncing the chain in a Raspberry Pi 4.}
        \label{fig:raspi-mem}
    \endminipage\hfill
    \minipage{0.45\textwidth}%
        \includegraphics[width=\linewidth]{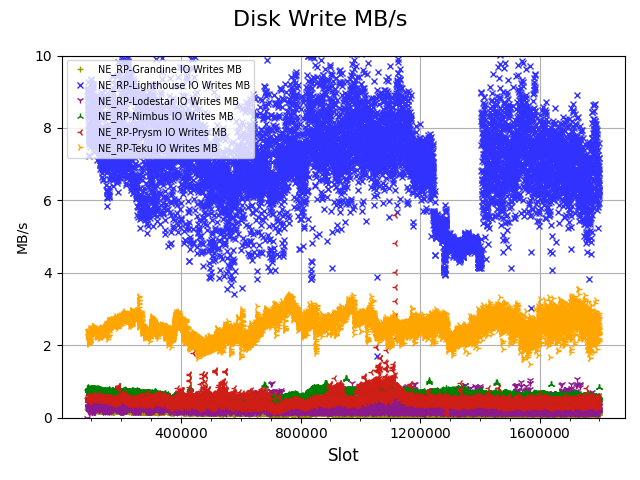}
        \caption{Disk writing speeds while syncing the chain in a Raspberry Pi 4.}
        \label{fig:raspi-disk-writes}
    \endminipage
\end{figure*}

A similar pattern was observed in terms of memory. Figure \ref{fig:raspi-mem} shows the memory allocation patterns from the CL clients in the Raspberry Pis. Teku keeps a steady performance by assigning $6GB$ of memory to the JVM. Prysm keeps allocating memory as \emph{Go} does not free memory unless the OS asks for it, reaching $5GB$ of memory in its latest synchronization stages. Lodestar, Grandine, and Nimbus remain with the lower profile at a lower limit of $3GB$, showing that not much memory is needed to sync up the chain. Finally, Lighthouse shows the same memory leak pattern but sooner this time, as the machine's total memory is reduced to $8GB$. We experienced three client crashes as more memory than available was asked. In this sense, each sudden drop in Figure \ref{fig:raspi-mem} belongs to each of the restarts of the client after the crash. As Figure \ref{fig:raspi-disk-writes} shows, with the shorter access to memory resources, the disk utilization remains steady from the beginning of the synchronization process as opposed to the previous hardware configurations. However, not all the clients have the same disk usage. In the figure, we can see that in synchrony with its memory leak, Lighthouse highly relies on disk write operations in comparison to other clients, multiplying by tree times the usage of Teku, and by more than eight times the rest.

\subsubsection{Performance}

\begin{figure*}[!htb]
    \minipage{0.45\textwidth}%
        \includegraphics[width=\linewidth]{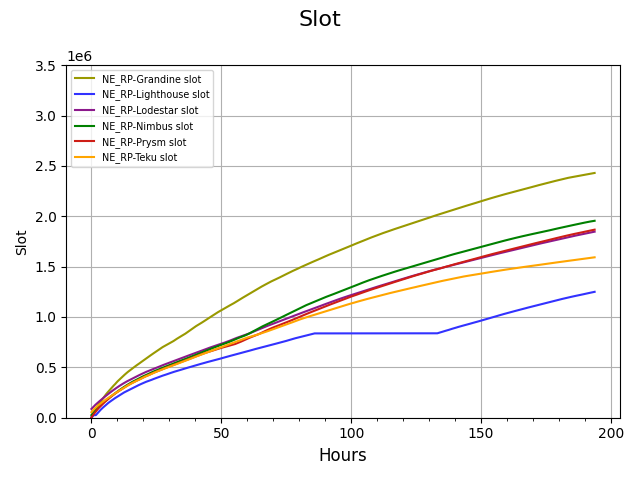}
        \caption{Synchronization speed of the chain in a Raspberry Pi 4.}
        \label{fig:raspi-sync-slots}
    \endminipage\hfill
    \minipage{0.47\textwidth}
        \includegraphics[width=\linewidth]{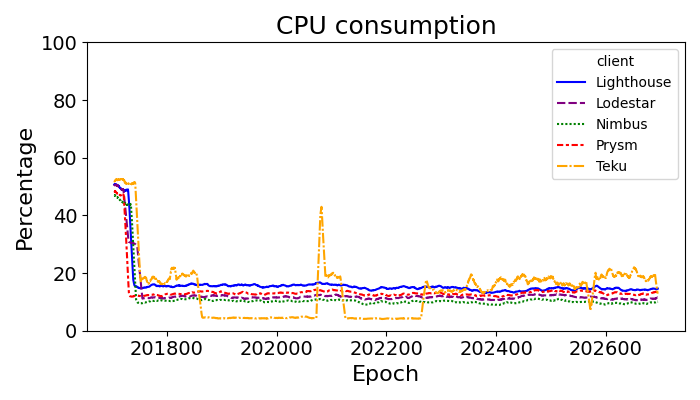}
        \caption{CPU usage by clients following the head of the chain.}
        \label{fig:regular-cpu}
    \endminipage
\end{figure*}

In the opposite direction as the one measured with the ``fat nodes", Figure  \ref{fig:raspi-sync-slots} shows decreasing the hardware resource clearly impacts syncing the chain. Making it clear, once again, that being slower while syncing the chain shouldn't be a determinant factor, as syncing via checkpoints significantly reduces the overall duration of this process. In any case, slower CPU cores, less memory, and slower reading and writing speeds from and to SD cards have drawbacks and impact performance. 

We have to take into account that the required hardware resources are just going to increase over time. Since the Beacon State is append only. As a result, CPU and memory requirements will increase with bigger state transitions to process. Even if the number of validators decreases, which doesn't seem feasible, at least in the short term, new updates to the protocol will always require more validations or tasks. Following this trend, after the merge, each consensus client node needs to be paired up with an execution client to follow up the head of the chain successfully. This ultimately means that the available resources have to be duplicated to run both parts of the system together.

Although there is a tendency to require faster and more extensive resources, we are definitely seeing a massive reduction when comparing it to the predecessor PoW consensus version. It is not crazy to think about having a \emph{Raspberry Pi} set up to participate in the network. However, having to purchase a faster SSD and probably extending to a secondary \emph{Raspberry Pi} to run the execution layer client (with its dedicated faster SSD) makes it, from our point of view, unattractive for most of the users. In this case, we suggest opting for a still low-profiled and low-powered device such as an \emph{Intel NUC} or similar.

\subsection{Regular performance}
\label{subsec:regular-performance}
Having the possibility to fully sync the chain from scratch and ensure that doing so is viable is a nice feature of the community. However, as we have introduced previously, current node deployments can benefit from syncing the chain using chain checkpoints. 

At any point, the resources taken from a client at a regular performance (following the head of the chain), are considerably smaller than those they need to sync up the chain. This subsection describes the resources that each client needs to perform after the synchronization phase.

To study a fairly more accurate representation of the actual consumption of the set of clients, we have also measured the resources that each of them needs to participate in the network while following the head of the chain.
To do so, we have decided to repeat the experiment with the Execution Client and the Consensus Client, both in the same machine. To promote a fair comparison between the CL clients, we paired all of them with a Nethermind node. The study was performed between the epochs 201699 and 202699, or in a human-readable format, between the 17th of May 2023 and the 21st of May 2023. The experiment was conducted in the Goerli network \cite{goerli}, using the client versions listed in Table \ref{tab:reg-perf-versions}. 

\begin{table}[]
    \centering
    \begin{tabular}{ccc}
        \hline
        Client & Version \\
        \hline  
        Prysm & 4.0.3 \\
        Lighthouse & 4.1.0 \\
        Teku & 23.4.0 \\
        Nimbus & 23.5.0 \\
        Lodestar & 1.8.0 \\
        \hline
    \end{tabular}
    \caption{Table with the versions used during the \emph{regular performance} study.}
    \label{tab:reg-perf-versions}
\end{table}

For this specific measurement, as we had to run one EL plus CL client pair on each machine, we discarded the option of using a low-resource machine such as the \emph{Raspberry Pi}. Just by the requirements/recommendations for both clients regarding Disk I/O operations and the required memory to operate, on single Raspi with 8GB of RAM and a micro SSD card are not enough. Thus, the experiments were done on a set of machines whose hardware is displayed in Table \ref{tab:regular-hardware}.

\begin{table}[]
    \centering
    \begin{tabular}{ccccc}
        \hline
        Provider & CPU & Memory & Storage \\
        \hline  
        OVH     & 8c.               & 32GB     & 900GB SSD \\
        \hline
    \end{tabular}
    \caption{Hardware configuration of the control machines used to measure the resources under a regular workload of the clients.}
    \label{tab:regular-hardware}
\end{table}

\subsubsection{CPU utilization} 
Unlike the chain synchronization process, the regular performance of both EL and CL clients doesn't require much CPU. Figure \ref{fig:regular-cpu} shows CPU usage by each of the combined clients as the percentage from the total available in the machine. The first slots in the figure, show that despite being syncing from a trusted node's checkpoint, the whole process of validating it and taking it up with the head of the chain is way more demanding than just following it up. The next epochs in the chart show that both CL and EL need less than 40\% of an eight-core machine to handle the propagation of blocks, attestations, and aggregations, as well as the validation of the same ones and recomputing the state. Bast reduction in contrast to its previous PoW consensus mechanics prior to the merge. 

There are a few more insights to take from the figure, though. Despite the difference being not much across the CL clients, we see that some are more efficient than others. Nimbus, Prysm, and Lodestar are the most efficient clients, taking between 11\% to 13\% of the CPU to perform all the consensus tasks. On the other hand, Teku and Lighthouse are the ones requiring more CPU, with Lighthouse keeping an 18\% of average CPU usage and Teku being the most unstable client fluctuating its CPU usage between 15\% and 23\%, sometimes reaching 30\% of use. 

We have to mention, though, that there were some infrastructure problems with Teku, who crashed two times during the experiment for a lack of space for the machine. These interruptions are clearly visible in the figure, wherewith sudden drops in Teku's CPU usage to a flat and stable 3\% to 7\% which we can attribute to Nethermind. These drops are also followed by spikes of 40\% to 60\%, which are related to the client catching up on the blocked missed while it was down.

\subsubsection{Memory utilization} 
More different memory usage patterns were measured on the machines during the study. In contrast to CPU usage, memory usage increases once the client follows the chain's head. Figure \ref{fig:regular-mem} shows that the client difference is more noticeable than the CPU one. Nimbus and Lighthouse are the most optimized clients with averages of (remember this is the aggregation of memory between Nethermind and the clients) 15GB and 18GB, respectively. Lodestar follows them quite closely, with an average of 19GB. Leaving Prysm and Teku as the highest memory-dependent clients with 20GB and 23GB of memory usage, respectively. 

Please note here that Teku is the only client that requires less memory following up the chain than syncing it up, and that part of the distribution shows the flat memory usage of Nethermind of 11GBs. 

These more updated measurements show that Lighthouse has fixed the memory leak and that even though the CPU is heavily used to sync up the chain, a bigger cache or more memory is more used by clients when performing under a normal state of the chain while following it. 

\subsubsection{Disk utilization}

\begin{figure*}[!htb]
    \minipage{0.47\textwidth}%
        \includegraphics[width=\linewidth]{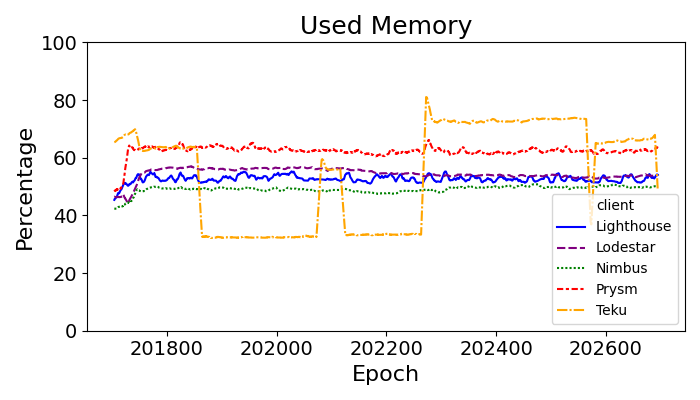}
        \caption{Memory usage by clients following the head of the chain.}
        \label{fig:regular-mem}
    \endminipage\hfill
    \minipage{0.47\textwidth}
        \includegraphics[width=\linewidth]{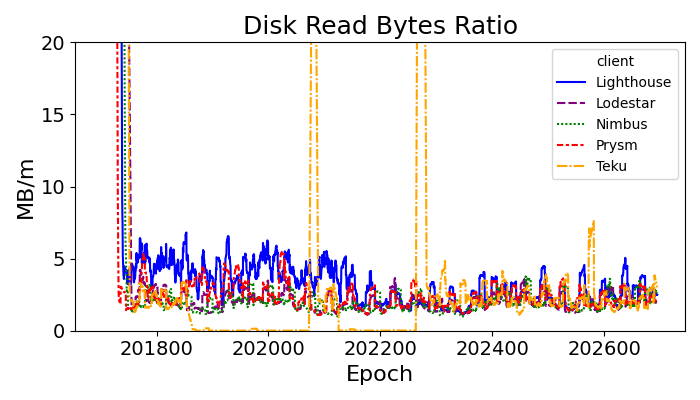}
        \caption{Disk reads by clients following the head of the chain.}
        \label{fig:regular-disk-read}
    \endminipage
\end{figure*}

From the perspective of the interaction with the disk, there are significant differences between clients. In contrast to the disk usage displayed in Section \ref{subsec:sync}, where clients could handle the persisting of blocks and states to DB in a more customized or optimized way, neither Figure \ref{fig:regular-disk-read} nor Figure \ref{fig:regular-disk-write} show clear differences in the number of Bytes read or written into the disk. Figure \ref{fig:regular-disk-read} shows a general 2MB per minute ratio of readings from the disk with some sporadic spikes from 4MB to 10 MB per minute. On the other hand, Figure \ref{fig:regular-disk-write} shows that writings are below 1MBs per minute, with more often spikes from 2MBs to 10MBs per minute.
This clearly shows that despite keeping more items and data in memory, the operation of a client requires way more reads and validations than actual writings to disk.

\subsubsection{Network bandwidth}
We've seen previously that the resources so far aren't anything out of the normal standard for computers. Meaning that the combo of EL and CL clients could easily run on modern pre-manufactured PCs or computers. However, can a normal router handle all the communication involved in the process?

\begin{figure*}[!htb]
    \minipage{0.47\textwidth}%
        \includegraphics[width=\linewidth]{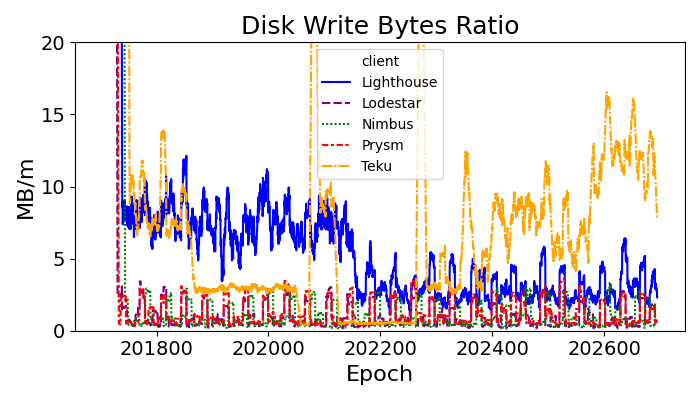}
        \caption{Disk writes by clients following the head of the chain.}
        \label{fig:regular-disk-write}
    \endminipage\hfill
    \minipage{0.47\textwidth}
        \includegraphics[width=\linewidth]{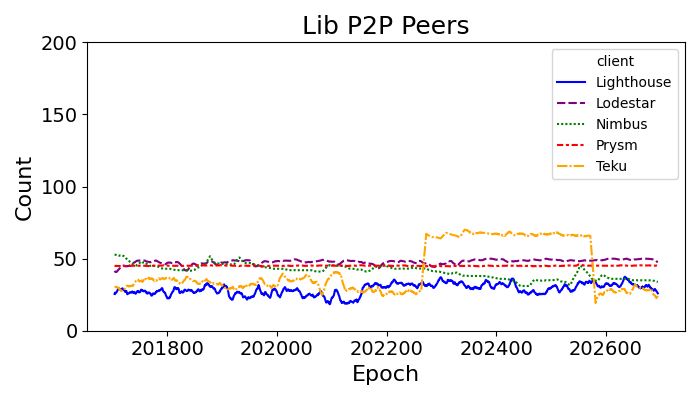}
        \caption{Concurrent node connections by the clients while following the head of the chain.}
        \label{fig:regular-peers}
    \endminipage
\end{figure*}

\begin{figure*}[!htb]    
    \minipage{0.47\textwidth}
        \includegraphics[width=\linewidth]{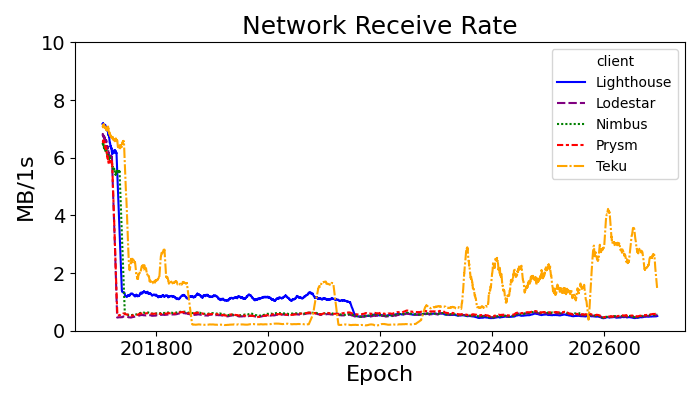}
        \caption{Network incoming bandwidth for the clients while following the head of the chain.}
        \label{fig:regular-net-in}
    \endminipage\hfill
    \minipage{0.47\textwidth}%
        \includegraphics[width=\linewidth]{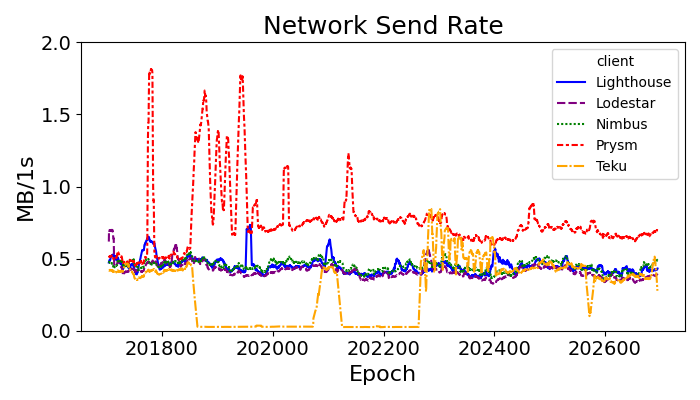}
        \caption{Network incoming bandwidth for the clients while following the head of the chain.}
        \label{fig:regular-net-out}
    \endminipage
\end{figure*}

The number of sent and received bytes over the network while following up the chain is directly proportional to the number of connections opened with other nodes in the network. The more nodes you connect to, the more messages or requests you receive and share. It doesn't matter if they are duplicated messages during the propagation of blocks or attestations; nodes will see an increase in network usage. 

Figure \ref{fig:regular-peers} shows each client's default target amount of peers. The figure shows that most of them, including Nimbus, Prysm, and Lodestar have a similar target of around 50 peers, which generally keep steady except for Nimbus, which sees a small drop of connected peers at the final epochs of the study. On the other hand, Teku and Lighthouse stay fluctuate more between the 30 to 40 concurrent connections. These numbers highly demonstrate that some clients were targeting a possible too-ambitious number of peers, like Nimbus, which went from 150 connections in \ref{fig:def-peers} at the beginning of its journey to 50 nowadays. 

To observe the impact of these concurrent connections in the downlink, we can use Figure \ref{fig:regular-net-in} which shows the amount of received MBs per minute for each of the client combos. Leaving the first synchronization from the checkpoint range, the figure shows that most clients, Nimbus, Prysm, and Lodestar, remain constant under one MB per minute mark. Clear contradiction to clients with more unstable peers (in terms of peer connections) like Teku and Lighthouse, which stay above the 1 MB mark, touching the received 2MBs per minute. 
This may look like a contradiction, since fewer peers send more traffic. We attribute this phenomenon to the fact that establishing a connection, which involves making the handshake and sharing certificates to ensure encryption between both partners, can be more demanding than just sharing messages. In this case, Teku is the chattiest client.  

Figure \ref{fig:regular-net-out} shows the impact in the uplink by providing the sent MBs per minute by the CL plus EL combo. In the figure, we can see how most of the client share a similar pattern, with Prysm being the chattiest client.

\subsection{CPU at slot time utilization}
\label{subsec:slot-cpu}

\begin{figure*}[!htb]
    \minipage{0.50\textwidth}
        \includegraphics[width=\linewidth]{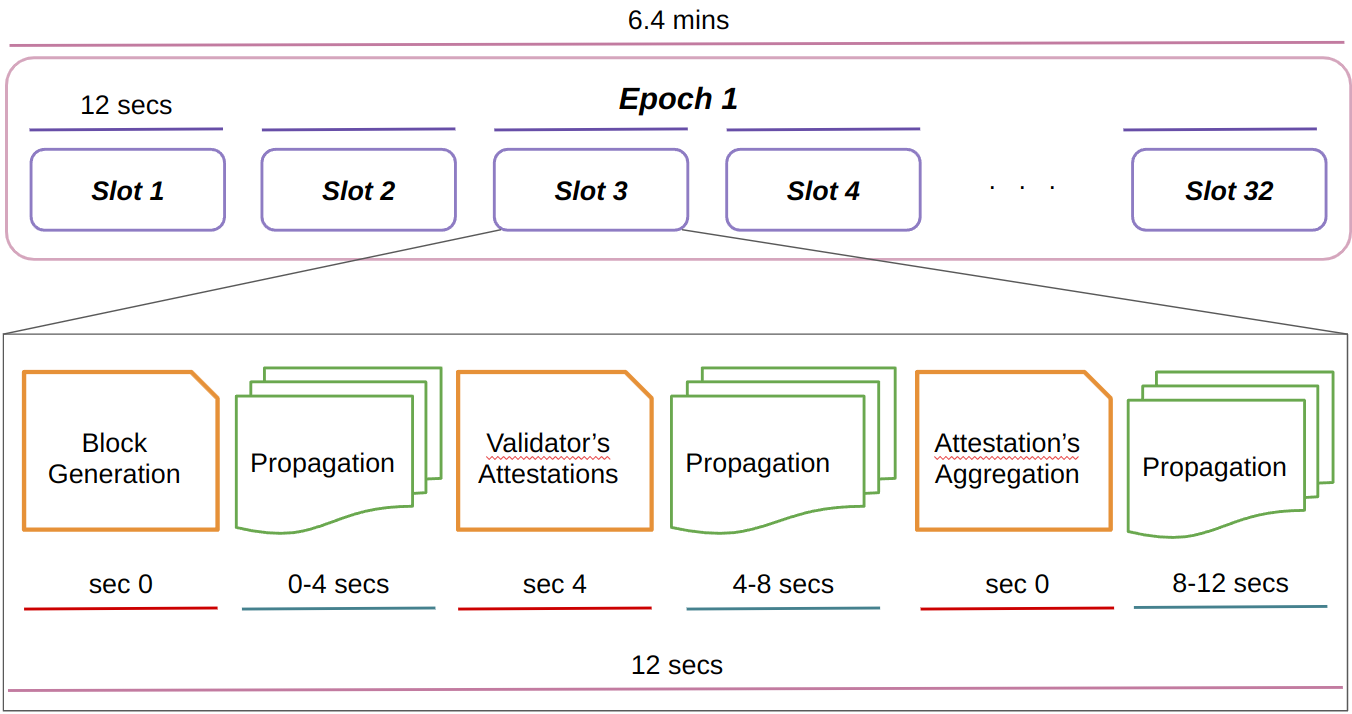}
        \caption{Description of the different validator task's time windows inside a single slot.}
        \label{fig:slot-time-description}
    \endminipage\hfill
    \minipage{0.47\textwidth}%
        \includegraphics[width=\linewidth]{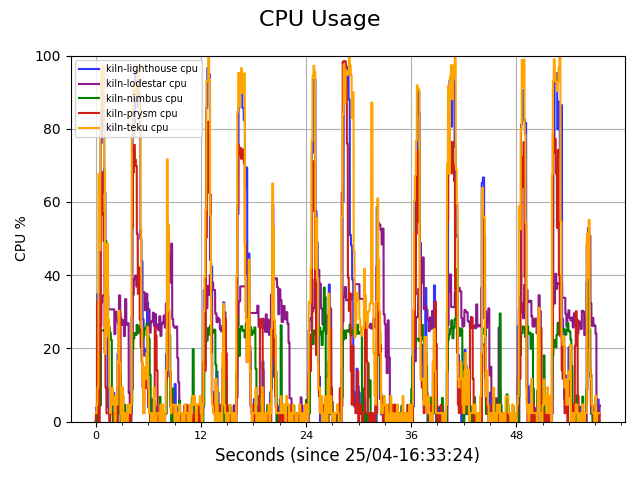}
        \caption{CPU utilization at the slot range.}
        \label{fig:cpu-util}
    \endminipage
\end{figure*}

We have previously introduced the overall resources needed by each of the clients while syncing and following the head of the chain. However, we haven't mentioned which processes trigger such intense use of CPU during the performance of a client. 
As we previously mentioned, Ethereum's PoS defines that the chain is organized in slots and epochs, where active validators split their duties across the epoch. This means that in every slot, a subset of active validators is in charge of receiving and validating the proposed block, having to then submit the attestation votes, and the aggregated attestations. Because there are three main duties to perform, each of the outcomes needs to be propagated over the peer-to-peer network layer. A detailed description of the expected time windows for each operation is described in Figure \ref{fig:slot-time-description}. In the figure, we can appreciate how each duty is followed by a 
$4$-seconds window to propagate each message over the network.

Of course, as the beacon validators are hosted by beacon nodes, the performance, validation, and broadcasting of these duties have 
an impact on the CPU workload of the nodes. Thus, we decided to measure the CPU workload of each machine with a bigger resolution, pairing it with their respective time inside the slot. Figure \ref{fig:cpu-util} shows the CPU workload of each client over five entire slots. Despite the workload of each client varies from the rest, the figure clearly shows three main spikes at around $4$ seconds from each other. The pattern is very clear when checking the CPU utilization by Prysm, Teku, and Lighthouse, where we can appreciate that:
\begin{itemize}
    \item The first spike corresponds to the arrival of the proposed beacon block. After one of the client's neighbors sends the block, this one not only has to validate the block's origin but also compute all the aggregated BLS signatures of each attestation aggregations on it, and update the existing beacon state with the new block. It is known that this process is generally CPU-intense since BLS signature aggregations require complex operations like elliptic curve pairings.
    \item The second spike belongs to the send and receive of the corresponding attestations. Attestations or votes are small messages. However, there are many of them, and the clients actively contribute to the validation and broadcasting of the messages that they see. This explains why, on some occasions, the second spike lasts longer than the first one.
    \item The last spike is the least clear one. It belongs to receiving and validating the aggregated attestations that will be included in the following beacon blocks. Due to their small size and the fewer number of messages that are sent, the last spike is the smallest one. 
\end{itemize}

We've seen that the CPU utilization can be easily mapped to each of the operations that the beacon node has to do. However, not all the nodes handle the workload in the same way. The behaviour of Prysm, Lighthouse, and Teku have the most visual explanations of the three spikes. Nimbus and Lodestar show different profiles. 
They suffer smaller CPU profiles but for longer periods. It is interesting to observe how Nimbus's spikes stay under 40\% of CPU utilization, keeping it for less than half a second compared with the three former clients. This smaller but longer CPU utilization means that Nimbus can handle the block arrival more efficiently, reducing the risk of not being able to process a block ``fast enough" if not enough resources are available by the hosting machine.   
Conversely, the Lodestar behaviour is the hardest to deduce from the charts. Its CPU is more used over a larger period, showing that it is either not that optimized to process a new block arrival or struggles when having to share back the block with the rest of the network.  


\subsection{Beacon API performance}
\label{subsec:beacon-api}

It is essential to mention that running an Ethereum node doesn't only concern users or entities that want to run validators on top of them. Accessing data from the blockchain is as important as reaching a consensus on the chain. Thus, many companies and researchers are actively participating in the network with the final goal of accessing, analyzing, or simply selling chain-related data. 

Most users access on-chain data such as transaction status, chain status, and validator performance through chain explorers such as Etherscan \cite{etherscan}, EthSeer \cite{ethseer} or Beaconcha.in \cite{beaconcha}. However, to support these web applications, they must interact with chain data through the REST API that most clients offer. There is a defined standard number of endpoints \cite{beacon-api} \cite{execution-api} that the clients are supposed to offer. However, the performance of retrieving that information from the API is not standard among the clients. 

Since quick access to this data might be essential for some users, we have bench-marked the performance of each client's APIs using their archival mode. Since not all the clients offer that mode, we have only focused on the ones that did, so the time comparison was fair for the clients. The benchmark was implemented by performing multiple queries at the same endpoint for different clients. The selected query was 
$/eth/v1/beacon/states/[slot]/validator\_balances?id=[validator\_id]$, 
which forces to recalculate a beacon state in the past returning the balance of a given validator. All the clients were asked the same queries in the same order, which were selected in a random order to ensure the recalculation of the past beacon state.    

\begin{figure*}[!htb]
    \centering
    \minipage{0.43\textwidth}
        \includegraphics[width=\linewidth]{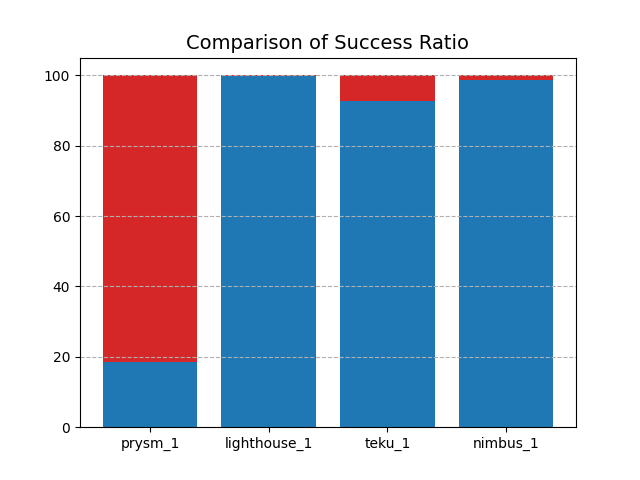}
        \caption{Beacon node's successful responses for a single concurrent request.}
        \label{fig:api-1-success}
    \endminipage\hfill
    \minipage{0.43\textwidth}%
        \includegraphics[width=\linewidth]{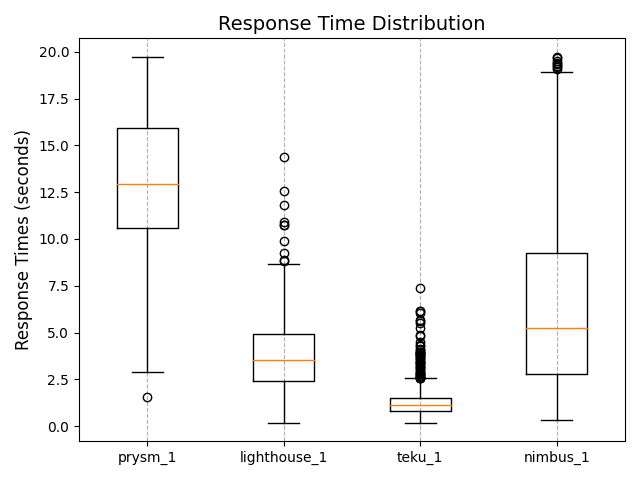}
        \caption{Beacon node's response times for a single concurrent request.}
        \label{fig:api-1-times}
    \endminipage
\end{figure*}

We performed two sets of experiments. The first one consisted of $1.000$ sequential queries performed with an in-between delay of $10$ seconds (for preventing the exhaustion of the resources of the beacon node). Figure \ref{fig:api-1-success} shows the success ratio of each client supporting the archival mode, where the low success of Prysm catches our attention. Digging into the possible root of the problem for such a low success rate, we discover that Prysm synced until the last slot we added to our query randomizer. However, by checking the response times in Figure \ref{fig:api-1-times}, the successful calls didn't perform that well compared with the rest of the clients. 
Prysm is known for being the only client that offers endpoints for both gRPCs and HTTP REST API calls. They strongly believe that gRPCs are better and faster. Even though this is a legit statement, at Prysm's beacon node level, the REST API calls are translated into gRPC on their arrival and vice-versa when returning the response, making the process slower than other clients. Given that the rest of the beacon nodes communicate with their validator client through the REST API, it leaves Prysm in a lower step towards client interoperability. Checking the rest of the time responses, we do remark that Teku managed to reply to all the queries in under $10$ seconds, with a median of $1.23$ seconds.

\begin{figure*}[!htb]
    \minipage{0.43\textwidth}
        \includegraphics[width=\linewidth]{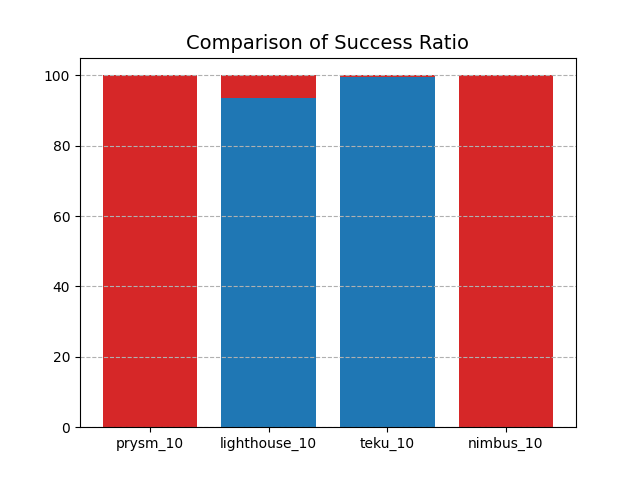}
        \caption{Beacon node's successful responses for ten concurrent requests.}
        \label{fig:api-10-success}
    \endminipage\hfill
    \minipage{0.43\textwidth}%
        \includegraphics[width=\linewidth]{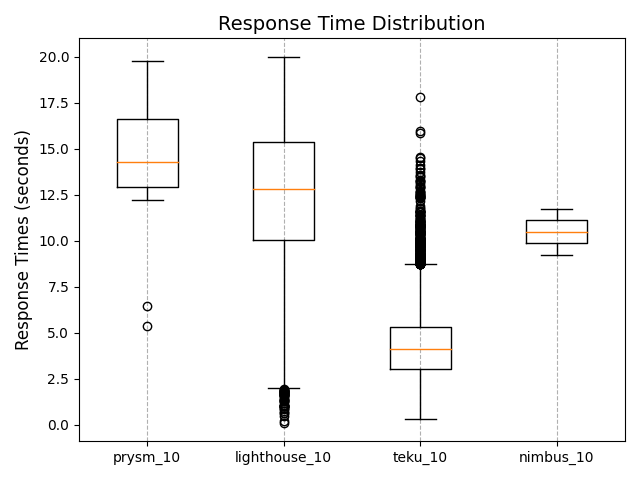}
        \caption{Beacon node's response times for ten concurrent requests.}
        \label{fig:api-10-times}
    \endminipage
\end{figure*}

In the second experiment, we increased the number of performed concurrent queries from one to ten, checking each client's limits. Figure \ref{fig:api-10-success} shows that only Teku and Lighthouse managed to keep the performance at such a level of demand, whereas Prysm and Nimbus stayed behind. Regarding the response times of the ten concurrent queries shown in Figure \ref{fig:api-10-times}, most of the response times distributions got higher and larger. Lighthose's and Prysm's response times got more concentrated around the predefined \emph{Timeout} of $20$ seconds. As expected, Lighthouse's median moved from around $3.1$ seconds to $13$ seconds, while Teku's median response time increased up to $4.2$ seconds. 

The clear difference between the client's throughput and response times is easily attributable to the different end-user targets. While Prysm and Nimbus might be performance-focused, Lighthouse and Teku have a larger business-research-oriented focus, where Lighthouse offers a larger set of API endpoints for data accessibility, and Teku was built with the idea of healing many requests coming from Infura \cite{infura}.

\subsection{Learning from the past}
\label{subsec:inestability}

Syncing up the historical records of the chains, generally known as blocks, might give more information than one would imagine. Back to the first stages of this study, when we were putting to test our methodology with the \emph{Medalla} testnet \cite{medalla-testnet} on the 2nd of November 2020 (prior even to the official launch of the Beacon Chain), we spotted several singularities across all the clients.

\begin{figure*}[!htb]
    \minipage{0.43\textwidth}
        \includegraphics[width=\linewidth]{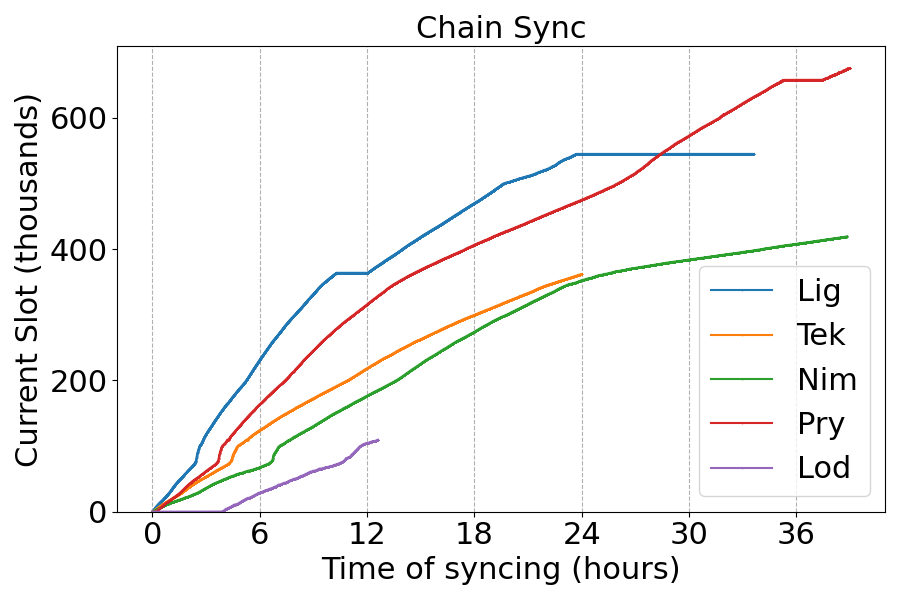}
        \caption{Chain synchronization time by clients in the medalla testnet.}
        \label{fig:past_sync_speed}
    \endminipage\hfill
    \minipage{0.43\textwidth}
        \includegraphics[width=\linewidth]{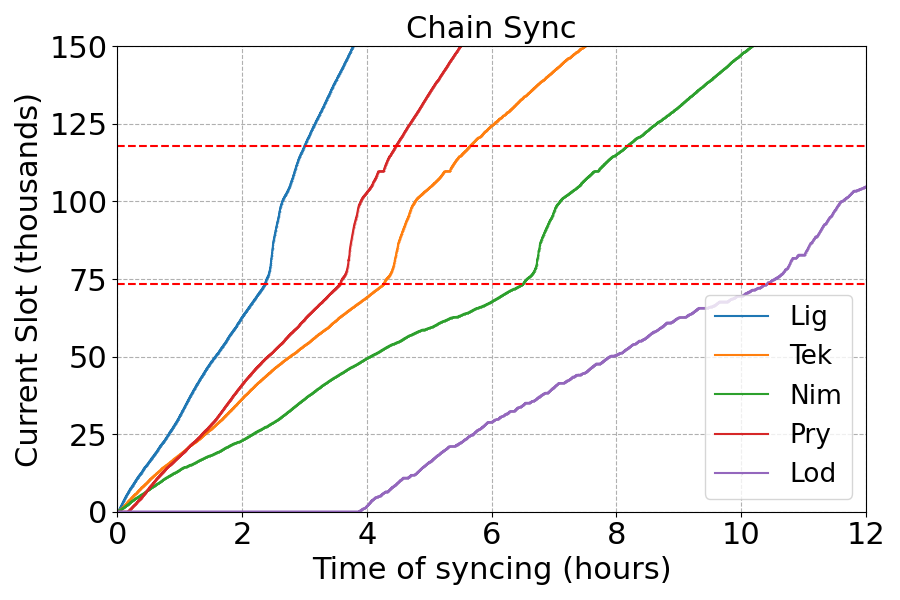}
        \caption{Zoomed slot synchronization speed of consensus clients.}
        \label{fig:past_sync_zoom}
    \endminipage
\end{figure*}

At the moment of syncing the testnet, the slot synchronization speed shown in Figure \ref{fig:past_sync_speed} didn't show anything relevant at first: a similar synchronization speed for most clients. At the same time, Lighthouse and Prysm were the fastest clients. But taking a closer look at the beginning of the syncing process, we could identify a small spike of synced slots across all the clients. 
Zooming into the range of slots 70,000 and 120,000, Figure \ref{fig:past_sync_zoom} showed that, in fact, the spike was constant across all the clients and not a singularity on only one of them. 

\begin{figure*}[!htb]
    \minipage{0.43\textwidth}%
        \includegraphics[width=\linewidth]{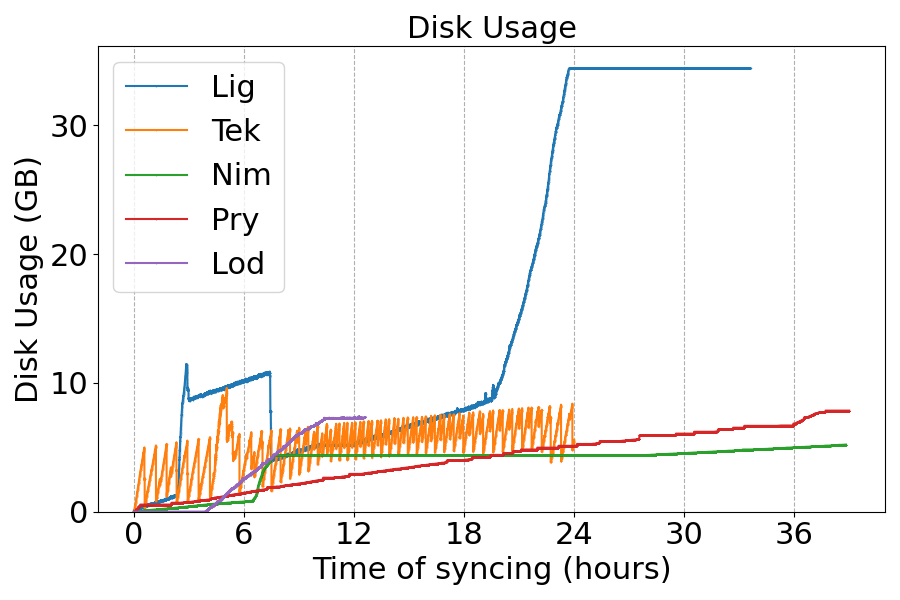}
        \caption{Disk usage of consensus clients during the medalla testnet synchronization.}
        \label{fig:past-disk}
    \endminipage\hfill
    \minipage{0.43\textwidth}%
        \includegraphics[width=\linewidth]{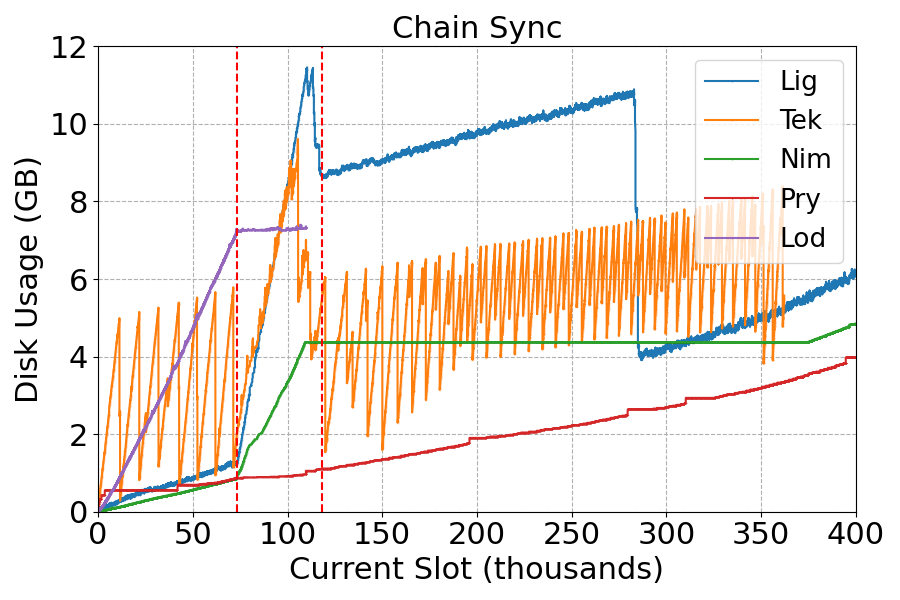}
        \caption{Disk Usage of Ethereum clients for slots.}
        \label{fig:past-disk-slot}
    \endminipage
\end{figure*}

Digging more into the anomaly, another metric suffered a similar distinction pattern around the same range of slots. Figure \ref{fig:past-disk} shows the disk usage measured by each client. The figure shows two clear spikes in two of the controlled clients, Lighthouse and Teku, around the 2nd and 7th hour of the synchronization process. 
In our attempt to correlate both anomalies, Figure \ref{fig:past-disk-slot} represents the disk usage using the synchronization slot as a reference in the X-axis. In the figure, the pattern gets more clear for all the clients except for Prysm, which barely notices any perturbance.  
There are several points to remark here: 
\begin{itemize}
    \item The 70.000 to 120.000 slot period corresponds with a non-finality period of the Medalla testnet which was caused by erroneous rough time responses witnessed by Prysm clients~\cite{medalla-nonfinality-august}.
    \item Figure \ref{fig:past-disk-slot} shows that the disk usage reaction of Teku, Lighthouse, and Nimbus is to spike the disk usage, while Lodestar offers the exact opposite reaction keeping the disk usage flat.
\end{itemize}

Considering the relatively similar behavior of Lighthouse and Teku, where the sharp increase of disk usage ends with a sharp drop, it is interesting to notice how Teku reduced the time of higher disk usage, while Lighthouse keeps running for several hours with additional data before dumping it. 
This is due to the dual-database system that Lighthouse and some other clients use: a \emph{hot} database that stores unfinalized beacon states, and a \emph{cold} database that stores finalized beacon states. As they sync through a large patch of non-finality, their hot databases grow large until they reach finality and then migrate this state into the cold database.

On the other hand, Nimbus's rise in disk storage is not as sharp as Teku and Lighthouse. However, it did not reduce its storage afterward (unlike Teku and Lighthouse). Oddly, we can notice that Lodestar's disk usage increases more rapidly than any other client until the start of this non-finality period, when it stops growing. 
Prysm's disk usage continues its trend without any variations as if it was not perturbed by the non-finality period. This is because Prysm clients only save finalized states every 2048 slots. This keeps disk utilization to a minimum. During non-finality, they do not save unfinalized states to disk, which allows them to prevent the database from unnecessarily growing.  
However, doing this comes at a cost, as they now keep everything in memory, if they need to retrieve a particular (unfinalized) state and it's been a while since finality, they have to regenerate it. Doing this puts a non-trivial amount of pressure on the CPU, making harder to
keep track of all the different forks.

\begin{figure*}[!htb]
    \minipage{0.43\textwidth}%
        \includegraphics[width=\linewidth]{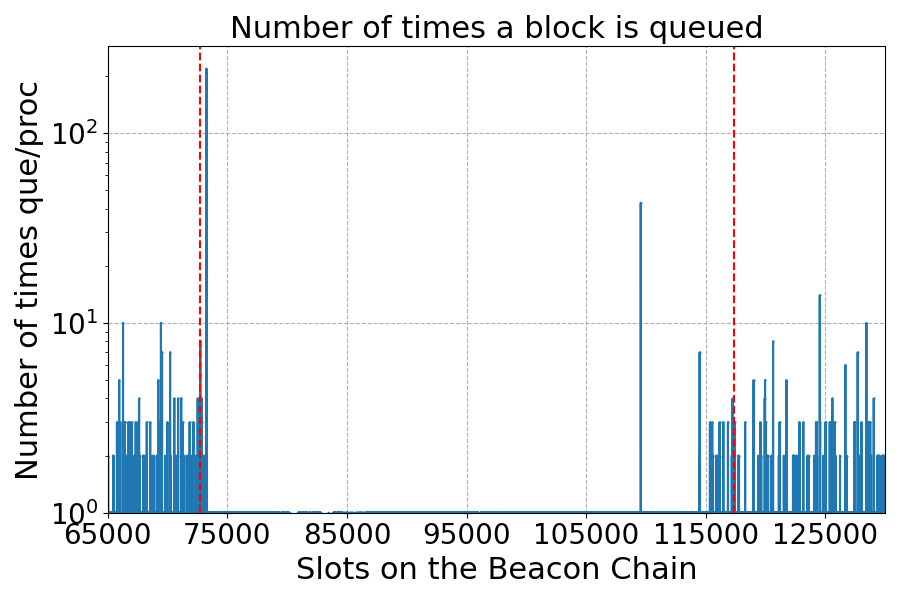}
        \caption{Number of times a block is queued to be persisted.}
        \label{fig:clients-slot}
    \endminipage\hfill
    \minipage{0.43\textwidth}%
        \includegraphics[width=\linewidth]{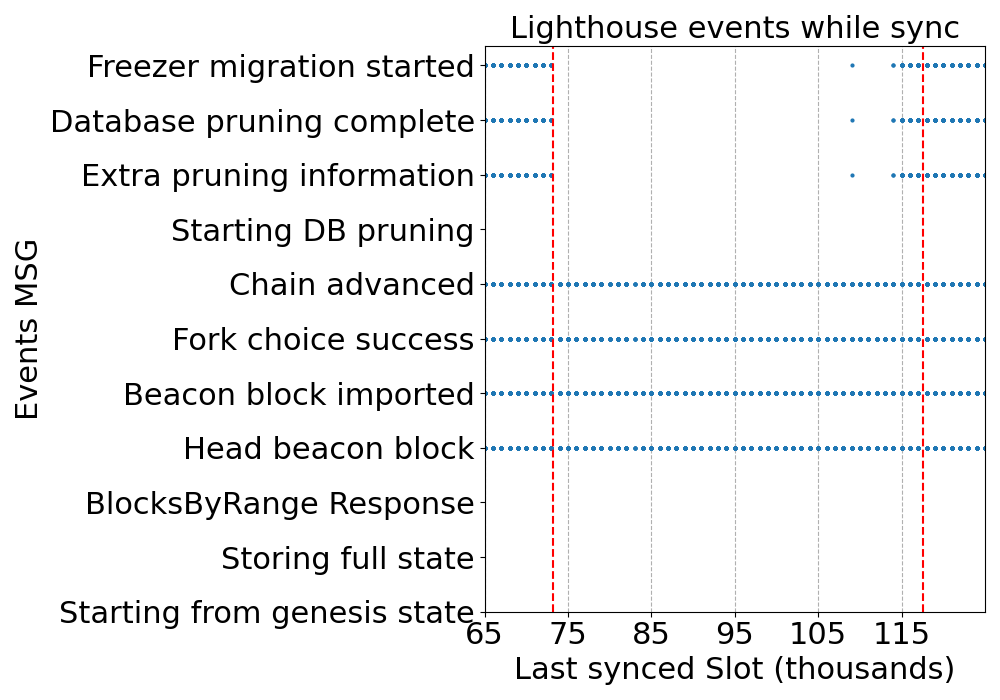}
        \caption{Events timeline for Lighthouse.}
        \label{fig:clients-current-slot-zoomed}
    \endminipage
\end{figure*}

The steeper syncing curve around slot $100,000$ previously seen could imply that during that time, there was little information to process (lots of missed blocks due to a lack of consensus). Therefore, clients can move faster in the syncing process. 
However, this does not seem to fit with the accelerating disk usage observed during the same period. 
To look deeper into this question, we used Lighthouse logs to analyze the number of times a block was queued and/or processed for each slot during the non-finality period. The results, depicted in Figure~\ref{fig:clients-slot}, show that during this period, there were almost no blocks queued, which seems to be consistent with the accelerated syncing speed. However, we also noticed that at the beginning of the non-finality period, at exactly slot $73,248,219$, there were blocks being queued (note the logarithmic Y axis), followed by a sudden drop of blocks to queue for more than $30,000$ slots. This clearly shows a considerable perturbation in the network.

We assume that the accelerating disk usage is related to an increase in the state stored in the client's database, which might be linked to the difficulty of pruning states during a non-finality period. Thus, to corroborate our hypothesis, we analyzed Lighthouse's detailed logs and plotted the frequency at which different events get executed. 
Figure~\ref{fig:clients-current-slot-zoomed} lists 11 different types of events. We can see that during the non-finality period, four types of events rarely get executed: Freezer migration started, Database pruning complete, Extra pruning information, and Starting database pruning. This demonstrates that the client could not prune the database during this period, which is consistent with the rapid disk usage increase.

Although multiple things remain to be understood about the behavior of some clients during a non-finality period, this paper demonstrates that it is possible to identify such a network disturbance by simply looking at the resource utilization of the clients.

\section{Discussion}
\label{sec:discussion}

In our journey through this exhaustive evaluation, we found multiple strong points as well as room for improvement in all Ethereum CL clients. Our objective with this study was fourfold: i) introduce the necessary hardware resources to run a complete Ethereum node; ii) dissect the necessary hardware resources on each of the stages of an Ethereum node (syncing vs 
 following the head); iii) introduce the different implementations available to participate in the network; iv) and last but not least, provide an overview of how the network can affect the available resources.

\subsection{Strengths and weaknesses of the clients}
The evaluation presented above gives empirical data about how the Ethereum CL clients perform under different hardware configurations and network scenarios. However, many other aspects also play an important role when choosing the software that will be deployed on an operational platform, such as documentation of the clients' usage, functionalities of the exposed API (important if the client will be used as an entry point to Ethereum's on-chain data). 
Although some of these aspects might be subjective, we try to cover some of those aspects together with the empirical data measured, discussing the strengths and points for improvements of each Ethereum CL client.

\subsubsection{Prysm}
Prysm has one of the best user experiences among its clients. It is easy to set up and deploy on its default configuration. The Prysmatic Labs \cite{prysmatic-labs} team has done a remarkable job simplifying the deployment for non-technical users. 
However, on the other hand, it has room for optimization in several aspects. The documentation portal could be improved; finding information on how to configure certain parameters is not that intuitive. The API offered by the client could be highly improved. The API is generally used as a communication point between the validator and the beacon node using gRPCs. Despite gRPC being a nice alternative to the standard HTTP endpoint APIs, Prysm has to comply with the standard HTTP API that allows interoperability with other validator clients. And because they rely on gRPC by default, the performance of the HTTP endpoints gets massively impacted as each HTTP request has to be translated to gRPCs. The result of this performance impact is a slower API that can't support multiple requests simultaneously. The synchronization of the client in an archival mode (storing the beacon state checkpoints with a very low frequency for faster access to them) is also slower when compared with other clients.

\subsubsection{Lighthouse}
Lighthouse from Sigma Prime \cite{sigp} was the client with the most complete API. It has all the CL Beacon node API standard implemented, and they have extended the endpoints with others that we found interesting from a research or data analysis point of view. On the other side, the client on the first measurements had a memory leak while syncing the chain from Genesis, it also has the highest disk requirements among all clients and seems to be among the most chatty clients. As with Prysm, the number of disk IO management should be reviewed, as other clients have shown that it can be considerably reduced.

\subsubsection{Teku}
Teku seems to be one of the most stable clients with very complete documentation, in which it is really easy to find any execution option and command line flag. It has a very competent archival node mode, as Consensys \cite{consensys} uses it to provide all the information through Infura \cite{infura}. However, generally, its synchronization is pretty slow, and when it comes to the archival node (which definitely needs to be synced from Genesis), it takes a lot more space than comparing it with others. On the other side, its standard client has the lowest storage needs, and the archival mode has the fastest API response time of all clients. 
It is also really important to note that setting up the JVM correctly to avoid memory issues can be tricky at the beginning, so it might take a few tries to properly set it up.

\subsubsection{Nimbus}
Nimbus from Status \cite{status} is the client with the lowest CPU and memory requirements across all platforms and the fastest syncing open-source client. It is clearly the client better suited to run on low-power devices, but it also performs well on more powerful servers. On the other hand, its compilation and deployment are not as user-friendly as other clients. Also, the fact that the Beacon node and Validator node run on the same executable could be viewed as a feature but also as a disadvantage, as sometimes it is useful to stop the Validator client while keeping the Beacon node alive.  

\subsubsection{Lodestar}
Lodestar from ChainSafe \cite{chainsafe} is one of the latest CL clients to join the race and it is commendable to see that the software supports most of the features that the other clients offer. Also, it shows a fairly low resource consumption. However, Lodestar is not always easy to compile and deploy (except when using Docker), there were multiple outdated instructions in the documentation. It is also the slowest client to sync from Genesis and it does not offer archival mode.

\subsubsection{Grandine}
Grandine was the fastest client to sync across all. It seems to have a great parallelization strategy that outperforms other clients while syncing from Genesis. However, it is not sure how much this speed can impact the performance after syncing. Despite there are still many features in beta, clearly, the biggest drawback of this client is that it has not been open-sourced yet.

\subsection{Hardware recomendations}
We have shown that the running combo CL plus EL clients is way more memory-demanding under standard conditions, which involves clients following up with the canonical head of a finalizing chain.
However, we do see that in the moments of having to re-sync some part of the chain, either because there was a chain re-organization when some blocks didn't get enough votes and had to be dropped, forcing to resync the last finalized checkpoint or simply because the client went down, the CPU spikes 3-4 times that usage.

Similar resource spikes are expected if the network is experiencing difficulties in finalizing, where more uncertain states and blocks must be kept in the disk and memory.
So having that extra hardware, CPU, memory, and faster and bigger disks can make the difference between recovering faster from this rare behavior of the chain or being penalized by it because the client struggles to recompute and sync up the correct head of the chain because CPU, memory, or disk are at their 100\% capacity, but it is insufficient.

This shouldn't be extrapolated to choose renting the ``fattest" or ``biggest" possible machine to run a node, as we support a middle ground where hardware should be slightly overestimated to satisfy those sudden need spikes. However, we do believe that the current requirements and performance of the clients exceed the hardware of domestic low-performance hardware devices such as a \emph{Raspberry Pi}. 
Of course, some tweaks and upgrades can be done in the hardware of Raspberry Pis, like extending the disk speeds with external SSDs, buying more powerful modifications or alternatives, upgrading it with extension packs, and so on, but compatibility and the community easily troubleshoot complications might leave this option to enthusiasts. 

At the time of writing this paper, we found a sweet spot  for the hardware on machines with 8 CPU cores, preferably 32GB of memory, and a fast 2TB SSD disk. 
This is a good result since many PCs can satisfy these requirements. There are solid low-powered devices such as laptops, Inte-NUCS, Mac Minis, and so on that could run an Ethereum node without any problems and the need to build a custom PC. On the other hand, there are custom solutions like DappNode machines that can help and guide less experienced or technical users to maintain their nodes, which summarizes the configuration and maintenance of the node in a few clicks. 


\subsection{The network can benefit from your client choices}
We have already presented and discussed the different requirements of the different available CL clients. However, they all have shown to be reliable, and there isn't much difference between them that makes any of them a clearer better choice. We believe each of them has its target users, but we have to encourage users to try them all out and choose the one they feel more comfortable with. Client diversity is an important aspect of the network's resilience and ultimately the chain. So exploring the least popular client choices can not only help the community but also surprise us with a better performance than the one we expected.    

Following the same line of recommendations, we highly encourage users to lose the fear of playing around with spawning their own nodes and to stake from home. These are very good practices that help the decentralization and resilience of the entire network. Furthermore, there is a broad literature, forums, communication channels, and a friendly community willing to support setting up or troubleshooting the spin-up of Ethereum nodes.

\section{Conclusion}
\label{sec:conclusion}

In this paper, we have shown multiple aspects of all Ethereum CL clients while tested under different conditions. We have exposed their strengths as well as discussed some points for improvement. After all these experiments, it seems clear that the different CL client teams have focused on different aspects, users, and use cases and they excel in different points.

Perhaps the most important conclusion that should be highlighted, is that all Ethereum CL clients run well on different hardware platforms and configurations. They showcase the strong software diversity that Ethereum has, and this is hard to find anywhere else in the blockchain ecosystem. Overall, our evaluation demonstrates that the efforts of all CL client implementation teams and researchers involved have pushed the Ethereum ecosystem one step closer to a more sustainable and scalable blockchain technology.

\section{Acknowledgements}
\label{sec:ack}

This work has been supported by the Lido Ecosystem Grant Organization (LEGO), the Spanish TCO-RISEBLOCK (PID2019-110224RB-I00) project, the Ethereum Foundation under the Research Grant FY21-0356, and Protocol Labs under its Ph.D. Fellowship Program FY22-P2P. We want to thank the client developer teams for their support in troubleshooting different encountered issues and their feedback on this work. Also, to Paristosh from the Ethereum Foundation, for his implication and support in the project, and Izzy, for his constructive feedback on this study.
\bibliographystyle{IEEEtran}
\bibliography{references}

\begin{IEEEbiography}[{\includegraphics[width=1in,height=1.25in,clip,keepaspectratio]{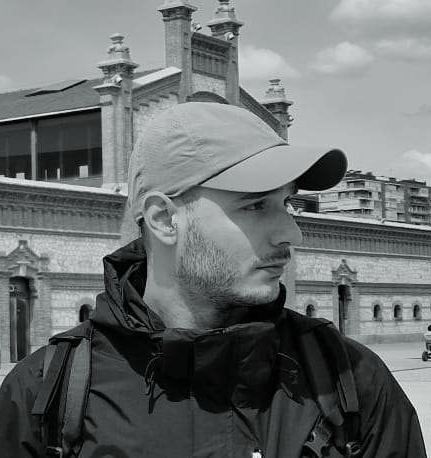}}]{First A. Author, Mr. Mikel Cortes-Goicoechea} was born in the Basque Country, Spain, in 1996. 
He graduated with a Bachelor's in Industrial Electronics and a Master's in Embedded Systems at the University of the Basque Country, Spain. After this, he became a Ph.D. candidate at the Polytechnic University of Catalunya while working as a Research Engineer at the Barcelona Supercomputing Center and MigaLabs, Spain. 
His research has focused on p2p networks, p2p protocols, and blockchain applications such as Ethereum, IPFS, and Filecoin.
He was awarded with a Ph.D. research fellowship from Protocol Labs, during which he could collaborate on various research projects with entities like Protocol Labs, the Filecoin Foundation, the Ethereum Foundation, the University of Cambridge, and the Codex Storage team at Status.
\end{IEEEbiography}

\begin{IEEEbiography}[{\includegraphics[width=1in,height=1.25in,clip,keepaspectratio]{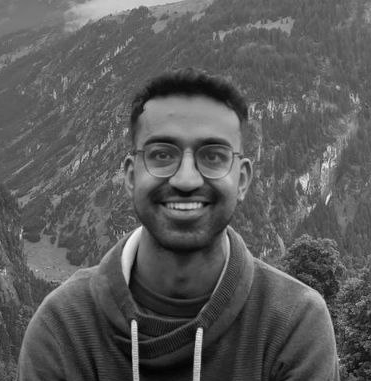}}]{Second B. Author, Mr. Tarun Mohandas-Daryanani} was born in Canary Islands, Spain in 1996.
He coursed Computer Science at the University of La Laguna (in the Canary Islands), and he continued his studies with his masters in Information Security Technology at the University Oberta in Catalunya. 
From 2018 to 2019, he participated in a scholarship with Telefonica, where he developed several tools to automate the data transfer between APIs and maintained some of the Video Platform infrastructure systems.
From 2019 to 2021, he was a cybersecurity consultant at Daimler Group Services Madrid, where his main role was reviewing the Software Development Lifecycle of the internal applications.
From 2021 onwards, he has been a Research Engineer at Migalabs, a small research group on the Ethereum blockchain, where he has developed both code and infrastructure. 
Among his main experiments and research, we can highlight some performance analyses: the resource analysis on the Ethereum Consensus clients and the Distributed Validator performance analysis in collaboration with the Obol team. Tarun is also the main maintainer of the GotEth tool, a lightweight beacon chain data extractor that powers the Ethseer website.
\end{IEEEbiography}

\begin{IEEEbiography}[{\includegraphics[width=1in,height=1.25in,clip,keepaspectratio]{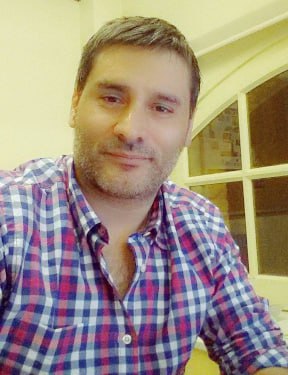}}]{Third C. Author, Dr. Jose L. Mu\~noz-Tapia}{\,} is a researcher of the Information Security Group, an associate professor of the Department of Network Engineering of UPC, and the director of the Master program in Blockchain technologies at UPC School. He holds a M.S. in Telecommunications Engineering (1999) and a PhD in Security Engineering (2003). His research interests include applied cryptography, network security, game theory models applied to networks and simulators, and distributed ledgers technologies.
\end{IEEEbiography}

\begin{IEEEbiography}[{\includegraphics[width=1in,height=1.25in,clip,keepaspectratio]{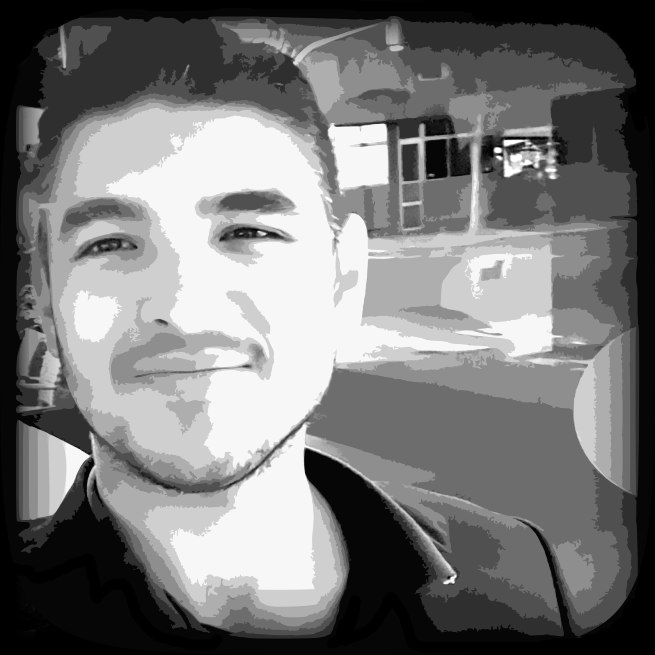}}]{Fourth D. Author, Dr. Leonardo Bautista-Gomez} is the founder and team leader of MigaLabs and he is also a senior researcher in the Codex team. He has been collaborating with the EF for more than 5 years and has received 5 research grants from the EF, plus several research grants from Lido, Obol and other institutions. He has over a decade of research experience in supercomputers, deep learning and blockchain technology. He has published more than 50 scientific articles and has received multiple international academic awards such as the IEEE TCSC Award for Excellence in Scalable Computing and the ACM/IEEE George Michael Memorial High Performance Computing Fellow. He did his Ph.D. at the Tokyo Institute of Technology and his Masters at the Pierre \& Marie Curie Paris 6 University.
\end{IEEEbiography}

\EOD

\end{document}